\documentclass[]{aa} 
\usepackage{natbib}
\usepackage{CJK}

\usepackage{graphicx}
\usepackage{txfonts}
\usepackage{xcolor}
\usepackage{multirow}
\def\h2{\ion{H}{II}}

\usepackage{lipsum}
\usepackage{tikz,xcolor}
\usepackage[colorlinks=true, allcolors=blue]{hyperref}
\definecolor{lime}{HTML}{A6CE39}
\DeclareRobustCommand{\orcidicon}{
        \begin{tikzpicture}
        \draw[lime, fill=lime] (0,0) 
        circle [radius=0.16] 
        node[white] {{\fontfamily{qag}\selectfont \tiny ID}};
        \draw[white, fill=white] (-0.0625,0.095) 
        circle [radius=0.007];
        \end{tikzpicture}
        \hspace{-2mm}
}
\foreach \x in {A, ..., Z}{\expandafter\xdef\csname orcid\x\endcsname{\noexpand\href{https://orcid.org/\csname orcidauthor\x\endcsname}
                        {\noexpand\orcidicon}}
}

\begin{document}

   \title{Discovery of widespread non-metastable ammonia masers in the Milky Way}


   \author{{\protect\begin{CJK*}{UTF8}{gkai}Y. T. Yan (闫耀庭)\protect\end{CJK*}} \inst{\ref{inst.mpifr}}\fnmsep\thanks{Member of the International Max Planck Research School (IMPRS) for Astronomy and Astrophysics at the universities of Bonn and Cologne.}\orcidB{}
    \and C. Henkel\inst{\ref{inst.mpifr},\ref{inst.xao}}\orcidA{}
    \and K. M. Menten\inst{\ref{inst.mpifr}}\orcidC{}
\and T. L. Wilson\inst{\ref{inst.mpifr}}
\and A. Wootten\inst{\ref{inst.nrao}}\orcidG{}
 \and {\protect\begin{CJK*}{UTF8}{gkai}Y. Gong (龚\protect\end{CJK*}\protect\begin{CJK*}{UTF8}{bkai}龑\protect\end{CJK*})} \inst{\ref{inst.mpifr}}\orcidD          {}
\and F. Wyrowski\inst{\ref{inst.mpifr}}
\and {\protect\begin{CJK*}{UTF8}{gkai}W. Yang (杨文锦)\protect\end{CJK*}}\inst{\ref{inst.nju},\ref{inst.mpifr}}\orcidY{}
\and A. Brunthaler\inst{\ref{inst.mpifr}}\orcidH{}
\and A. Kraus\inst{\ref{inst.mpifr}}
 \and B. Winkel\inst{\ref{inst.mpifr}}\orcidW{}
}

   \institute{
\label{inst.mpifr}Max-Planck-Institut f\"{u}r Radioastronomie, Auf dem H\"{u}gel 69, 53121 Bonn, Germany\\  \email{yyan@mpifr-bonn.mpg.de, astrotingyan@gmail.com}
\and\label{inst.xao}Xinjiang Astronomical Observatory, Chinese Academy of Sciences, 830011 Urumqi, PR China
\and\label{inst.nrao}National Radio Astronomy Observatory, 520 Edgemont Road, Charlottesville, VA 22903-2475, USA
\and\label{inst.nju}School of Astronomy \& Space Science, Nanjing University, 163 Xianlin Avenue, Nanjing 210023, People's Republic of China
}

   \date{Received XXX; accepted YYY}

 
  \abstract
  {We present the results of a search for ammonia maser emission in 119 Galactic high-mass star-forming regions (HMSFRs) known to host 22\,GHz H$_2$O maser emission. Our survey has led to the discovery of non-metastable NH$_3$ inversion line masers toward 14 of these sources. This doubles the number of known non-metastable ammonia masers in our Galaxy, including nine new very high excitation ($J,K$)~=~(9,6) maser sources. These maser lines, including NH$_3$ (5,4), (6,4), (6,5), (7,6), (8,6), (9,6), (9,8), (10,8), and (11,9), arise from energy levels of 342 K, 513 K, 465 K, 606 K, 834 K, 1090 K, 942 K, 1226 K, and 1449 K above the ground state. Additionally, we tentatively report a new metastable NH$_3$ (3,3) maser in G048.49 and an NH$_3$ (7,7) maser in G029.95. Our observations reveal that all of the newly detected NH$_3$ maser lines exhibit either blueshifted or redshifted velocities with respect to the source systemic velocities. Among the non-metastable ammonia maser lines, larger velocity distributions, offset from the source systemic velocities, are found in the ortho-NH$_3$ ($K=3n$) than in the para-NH$_3$ ($K\neq3n$) transitions.}


   \maketitle
%

\section{Introduction}
\label{introduction}

The first maser was obtained from a source of ammonia (NH$_3$) molecules by Charles H. Townes and his group members in the laboratory \citep{PhysRev.95.282,PhysRev.99.1264}. Various maser species were discovered in the interstellar medium (ISM), such as hydroxyl (OH) \citep{1965Natur.208...29W}, water (H$_2$O) \citep{1969Natur.221..626C}, and methanol (CH$_3$OH) \citep{1971ApJ...168L.101B}. Although thermal emission from NH$_3$ was discovered in 1968 by \citeauthor{1968PhRvL..21.1701C}, the first NH$_3$ maser in the ISM was detected 14 years later, in the $(J,K)$~=~(3,3) metastable ($J=K$) line towards the massive star-forming region W33 \citep{1982A&A...110L..20W}. The first highly excited non-metastable ($J>K$) ammonia masers were detected by \citet{1986ApJ...300L..79M} in the $(J,K)$~=~(9,6) and (6,3) lines. So far, a total of 34 NH$_3$ inversion transitions ($\Delta K$~=~0 and $\Delta J$~=~0) have been identified as masers in the ISM \citep[see][and references therein]{2022A&A...659A...5Y}. Even in its rare isotopolog $^{15}$NH$_3$, maser emission was detected in the (3,3) \citep{1986A&A...160L..13M}, (4,3) and (4,4) transitions \citep{1991A&A...247..516S} but only towards the high-mass star-forming region (HMSFR) NGC 7538. The ammonia transitions identified as masers in the ISM are summarised in Table~\ref{table_nh3-masers}. 

Ammonia masers are rare in the ISM compared to other maser species, i.e., those of OH, H$_2$O, and CH$_3$OH. Over the last five decades after the first detection of astronomical masers, numerous successful maser surveys were carried out in different molecules and led to thousands of detections in the Milky Way. They targeted, for example, OH masers \citep[e.g.,][]{1989A&AS...78..399T,1994ApJS...93..549L,1994A&AS..103..301H,1997A&AS..122...79S,1998MNRAS.297..215C,2012A&A...537A...5W,2019A&A...628A..90B}, CH$_3$OH masers \citep[e.g.,][]{1991ApJ...380L..75M,1993MNRAS.260..425C,2009A&A...507.1117X,2009PASA...26..454C,2016ApJ...833...18H,2016MNRAS.459.4066B,2017ApJ...846..160Y,2019ApJS..241...18Y,2019ApJS..244...35L,2022A&A...666A..59N,2020ApJS..248...18Y,2023A&A...675A.112Y}, and H$_2$O masers \citep[e.g.,][]{1979A&A....72..234G,1988A&AS...76..445C,1990ApJ...350L..41M,1993A&AS...98..589W,1993A&AS..101..153P,2006ApJ...651L.125W,2011MNRAS.418.1689U,2011MNRAS.416..178B,2011MNRAS.417..238M,2014MNRAS.442.2240W,2015MNRAS.453.4203X,2016MNRAS.459..157T,2016ApJ...822...59S,2018ApJS..236...31K,2022ApJS..261...14L}. All of these are collected and can be easily accessed from the online database, Maserdb\footnote{https://maserdb.net/} \citep{2019RAA....19...34S,2019AJ....158..233L,2022AJ....163..124L}.

So far, ammonia maser lines have only been detected in 32 sources. Among them, metastable NH$_3$ masers are quite common and have been detected in 22 different regions. Non-metastable ($J>K$) ammonia masers have been found in 14 objects \citep{2022A&A...666L..15Y,2022A&A...659A...5Y}. Only four sources host both metastable and non-metastable NH$_3$ masers. These are the HMSFRs DR 21 \citep{1983A&A...124..322G,1986ApJ...300L..79M,1994ApJ...428L..33M,1996ApJ...457L..47G}, W51 \citep{1986ApJ...300L..79M,1987A&A...173..352M,1995ApJ...450L..63Z,2013A&A...549A..90H}, NGC 6334 \citep{1995ApJ...439L...9K,2007A&A...466..989B,2007MNRAS.382L..35W}, and Sgr B2(M) \citep{2018ApJ...869L..14M,2022A&A...666L..15Y}. In Table~\ref{table_nh3-sources}, we summarise the sources that are known to host ammonia masers. The metastable NH$_3$ (3,3) masers are thought to be collisionally pumped \citep[e.g.,][]{1983A&A...122..164W,1990MNRAS.244P...4F,1994ApJ...428L..33M,1995ApJ...450L..63Z,1999ApJ...527L.117Z,2016ApJ...826..189M}. Pumping scenarios of other NH$_3$ transitions are still speculative. High angular resolution data show that the excitation of non-metastable NH$_3$ (9,6) masers in W51, Cepheus A, G34.26$+$0.15, and the Sgr B2 complex may be related to shocks by outflows or by the expansion of ultracompact (UC) \h2 regions \citep[][]{1991ApJ...373L..13P,2022A&A...659A...5Y,2022A&A...666L..15Y}. Furthermore, the NH$_{3}$ (9,6) maser stands out as being the strongest and most variable one in W51-IRS2. Its variability is comparable to the H$_2$O masers in the region \citep[e.g.,][]{2013A&A...549A..90H}.

There exists no systematic survey to search for ammonia masers so far. Therefore, we selected 119 HMSFRs with high NH$_3$ column densities ($N_{\rm NH_3}$ $\ge$ 10$^{15.5}$ cm$^{-2}$) that are known to host water masers from previous K-band surveys. The sample is mainly based on the APEX Telescope Large Area Survey of the Galaxy (ATLASGAL, \citealt{2009A&A...504..415S}) catalogs \citep{2012A&A...544A.146W,2018A&A...609A.125W}, and the Red MSX Source survey \citep[RMS,][]{2011MNRAS.418.1689U}. Our sample is listed in Table~\ref{table_sources} together with the systemic local standard of rest (LSR) velocities and the source beam averaged NH$_3$ column densities, which are based on ammonia (1,1), (2,2), and (3,3) thermal emission in previous studies with typical beam sizes of order 32" at Green Bank and 38" at Effelsberg. We performed a K-band line survey with the 100-m Effelsberg telescope of this source sample with the motivations (1): to search for NH$_3$ maser lines, and (2): to search for the higher metastable ammonia transitions, i.e. the NH$_3$ (4,4) to (7,7) lines with excitation levels up to 500 K above the ground state to reveal so far poorly studied warm gas components in the HMSFR sources. We also obtained H$_2$O and CH$_3$OH maser spectra simultaneously as well as quasi-thermal lines from other molecules. The data allow us to perform a new K-band spectral classification of massive star-forming clumps, uniquely complementing other surveys.

In this letter, we report the discovery of numerous non-metastable NH$_3$ and two probable metastable NH$_3$ maser sources in the Milky Way. Results for the high excitation metastable ammonia thermal transitions, the non-metastable ammonia thermal lines, as well as H$_2$O and CH$_3$OH maser spectra will be published in future papers.

\section{Observations and data reduction}
\label{observations}

The K-band line survey was performed with the 100-meter Effelsberg telescope\footnote{Based on observations with the 100-meter telescope of the MPIfR (Max-Planck-Institut f\"{u}r Radioastronomie) at Effelsberg.} in 2022 November, in 2023 February, April, May, and July. The S14mm double beam secondary focus receiver was employed to simultaneously cover the entire K-band frequency range, i.e. 18.0--26.0 GHz. The receiver band was divided into four 2.5 GHz-wide subbands with the frequency ranges of 18.0--20.5 GHz, 19.9--22.4 GHz, 21.6--24.1 GHz, and 23.5--26.0 GHz. Each subband has 65536 channels, providing a channel width of 38.1 kHz, changing from $\sim$0.62~km~s$^{-1}$ at 18.5 GHz to $\sim$0.44~km~s$^{-1}$ at 26.0 GHz. The observations were performed in position switching mode with the off position 10$\arcmin$ in azimuth away from the source. The half power beam width (HPBW) was 49 $\times$ 18.5/$\nu$(GHz) arcseconds, that is 49$\arcsec$ at 18.5 GHz, the frequency of the NH$_3$ (9,6) line. A high spectral resolution backend with 65536 channels and a bandwidth of 300 MHz was employed to measure the new NH$_3$ (9,6) maser sources, providing a channel width of 0.07~km s$^{-1}$ at 18.5 GHz. Pointing and focus calibrations were done at the beginning of the observations, during sunset and sunrise, as well as every two hours towards NGC~7027. The calibrator was measured between elevations of 30 and 60 degrees. Pointing was checked using nearby quasars prior to on-source integrations. The elevations of our targets were in a range of 15 to 55 degrees. The system temperatures were 60--130 K on a main-beam brightness temperature, $T_{\rm MB}$, scale. 

We used the GILDAS/CLASS\footnote{https://www.iram.fr/IRAMFR/GILDAS/} package \citep{2005sf2a.conf..721P} to reduce the spectral line data. The data were split into small frequency intervals with bandwidths of 300 MHz and were calibrated based on continuum cross scans of NGC 7027 \citep{2012A&A...540A.140W}, whose flux density was adopted from \citet{1994A&A...284..331O}. The $T_{\rm MB}/S$ ratios are 1.95 K/Jy, 1.73 K/Jy, and 1.68 K/Jy at 18.5 GHz, 22.2 GHz, and 24.0 GHz, respectively. Calibration uncertainties were estimated to be $\pm10$\%. All of the NH$_3$ lines covered in our observations are measured simultaneously, which ensures a good relative calibration.

\section{Results and discussion}
\label{results}

We detected 15 new ammonia maser sources, resulting in a detection rate of $\sim$13\%. These maser lines, including NH$_3$ $(J,K)$~=~(5,4), (6,4), (6,5), (7,7), (7,6), (8,6), (9,6), (9,8), (10,8), (11,9) and possibly the (3,3) line arise from energy levels of 342 K, 513 K, 465 K, 537 K, 606 K, 834 K, 1090 K, 942 K, 1226 K, 1449 K and 122 K above the ground state. The maser line parameters obtained by Gaussian fits are listed in Table~\ref{table_masers}. The masers are identified through three different ways: (1) narrow line widths compared to those of ammonia $(J,K)$~=~(1,1) thermal emission, (2) blueshifted or redshifted velocities with respect to the source's systemic LSR velocities, and (3) flux density variations. The first method can easily be verified by comparing the widths of putative maser lines with simultaneously observed quasi-thermal transitions.

\begin{figure*}[h]
\center
    \includegraphics[width=500pt]{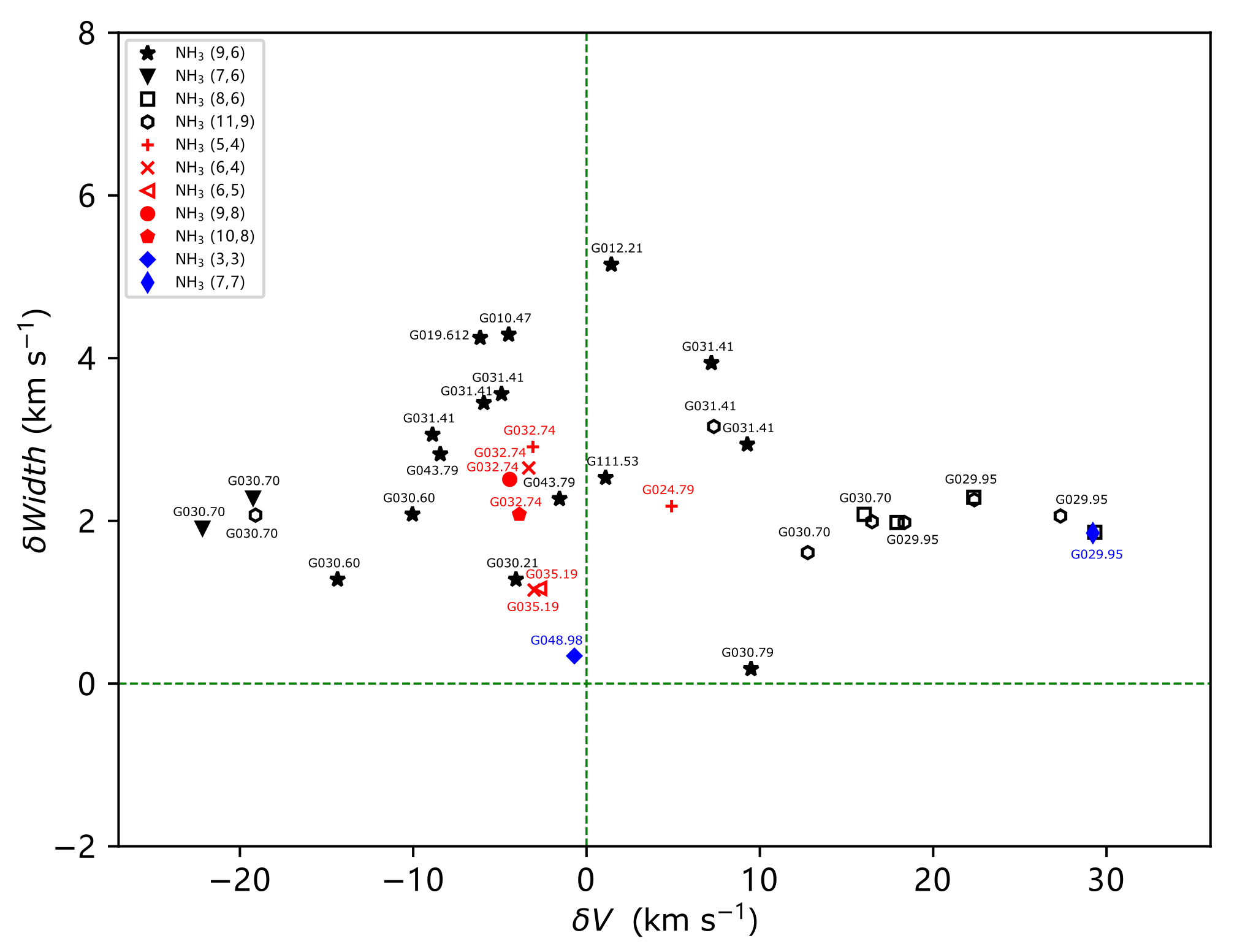}
  \caption{\label{comparison-hfs} Comparison of velocities and line widths of ammonia maser lines to the intrinsic line widths of ammonia $(J,K)$~=~(1,1) thermal emission. The number of data points is substantially larger than the number of newly detected sources due to the occasional presence of more than one maser line in a given source and more than one velocity component in a specific maser line. $\delta V$ = $V_{\rm LSR}$(maser)-$V_{\rm LSR}$(NH$_3$(1,1)), is the deviation from velocities of the NH$_3$ maser lines to those of NH$_3$ (1,1) thermal emission. $\delta Width$ = $\Delta V_{1/2}$(NH$_3$(1,1))-$\Delta V_{1/2}$(maser), refers to the difference between the line widths of ammonia maser lines and the intrinsic line widths of an individual NH$_3$ hyperfine structure component. The green dashed lines indicate positions with zero deviation. The metastable NH$_3$ transitions are marked in blue. Among the non-metastable NH$_3$ lines, ortho-NH$_3$ is presented in black, and para-NH$_3$ is given in red.}
\end{figure*}

Figure~\ref{comparison-hfs} shows the comparison of line widths of ammonia maser lines to the intrinsic line widths of hyperfine components of the $(J,K)$~=~(1,1) thermal emission as well as the difference between the systemic velocities of the sources and the maser velocities. We derived the intrinsic line width by using the hyperfine fitting in CLASS for the NH$_3$ (1,1) line. The line profiles of the NH$_3$ (1,1) thermal emission and maser transitions are presented in Fig.~\ref{all-11}. All of these maser lines have narrower features than the NH$_3$ (1,1) thermal emission. This further confirms their maser nature. Furthermore, their velocities are shifted with respect to the source's systemic LSR velocities, by at least 0.7 km~s$^{-1}$, and reaching up to 30 km~s$^{-1}$. This is similar to recent discoveries of non-metastable NH$_3$ masers with $\delta V\sim$ 10 km~s$^{-1}$ in Cep A, $\delta V\sim$ 4 km~s$^{-1}$ in G34.26+0.15, and $\delta V$ in a range of 0.3 km~s$^{-1}$ to 24 km~s$^{-1}$ toward the Sgr B2 complex \citep{2022A&A...659A...5Y,2022A&A...666L..15Y}. Fourteen of the new ammonia maser sources contain non-metastable ammonia masers, which doubles the number of non-metastable ammonia maser detections in our Galaxy. Among the non-metastable ammonia maser lines, larger velocity distributions are found in the ortho-NH$_3$ ($K=3n$) than in the para-NH$_3$ ($K\neq3n$) transitions. The velocity range of para-NH$_3$ masers is limited within $\pm$5~km~s$^{-1}$ with respect to the source's systemic velocities, marked as red in Fig.~\ref{comparison-hfs}. This is enlarged to about $\pm$30~km~s$^{-1}$ for ortho-NH$_3$ masers.

\begin{figure*}[h]
\center
    \includegraphics[width=400pt]{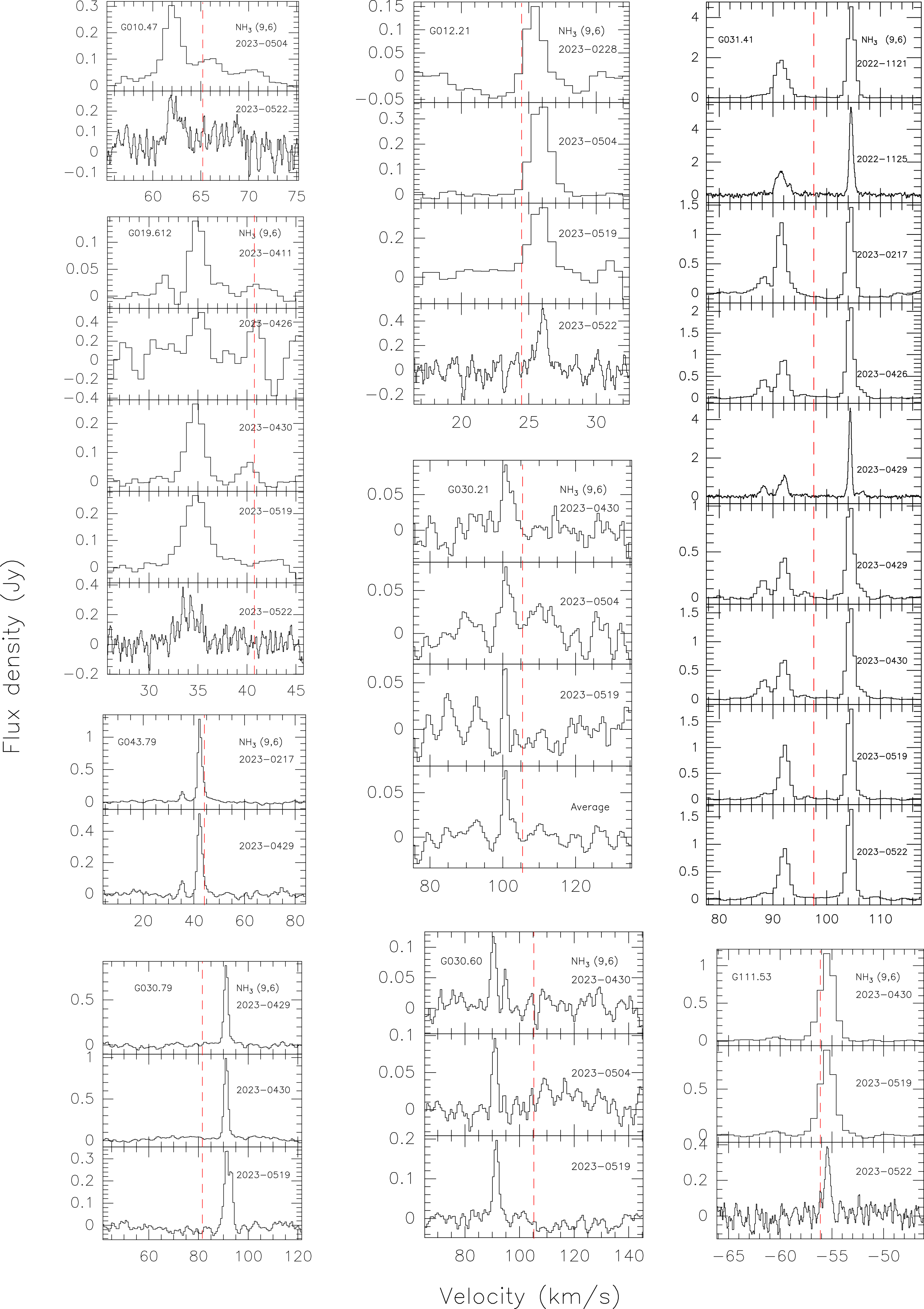}
  \caption{\label{M96} NH$_3$ (9,6) spectra observed at different epochs towards nine sources. The dashed red lines indicate the systemic velocities of the sources.}
\end{figure*}

Among the 14 non-metastable ammonia maser sources, NH$_3$ (9,6) masers are the most common and are detected in nine objects. The spectra of NH$_3$ (9,6) masers are shown in Fig.~\ref{M96}. The observations at different epochs indicate that the flux densities of these NH$_3$ (9,6) masers vary by at least 50\% over time scales of several months. Even within two days, variations were observed in G031.41+0.30 (hereafter G031.41), similar to that detected by \citet{2013A&A...549A..90H} towards W51-IRS2. The exception is G030.21-0.19 (hereafter G030.21): its NH$_3$ (9,6) flux density stays constant for 20 days. In order to increase the signal-to-noise ratios (S/N) of NH$_3$ (9,6) spectra towards G030.21, we average all three measurements at different epochs. The spectra and fitting results are presented in Fig.~\ref{M96} and in Table~\ref{maser_fitting}, respectively. Five targets, G010.47+0.03 (hereafter G010.47), G012.21-0.10 (hereafter G012.21), G019.61-0.23 (hereafter G019.612), G031.41, and G111.53+0.76 (hereafter G111.53), were also observed in the high spectral resolution mode (Fig.~\ref{M96}). These data show that the NH$_3$ (9,6) masers contain narrow components with line widths smaller than 1.0 km~s$^{-1}$ and some even below the channel width of the wide band spectra (Sect.~\ref{observations}), i.e. smaller than 0.6 km~s$^{-1}$. In eight sources, G010.47, G012.21, G019.612, G030.21, G030.60+0.18 (hereafter G030.60), G030.79+0.20 (hereafter G030.79), G043.79-0.13 (hereafter G043.79), and G111.53, we detect only NH$_3$ (9,6) masers. Five of these eight targets show only blueshifted (9,6) features, and the other three only have redshifted features. G031.41, the ninth object, is unique in that it hosts both blueshifted and redshifted NH$_3$ (9,6) maser components.

\begin{figure}[h]
\center
    \includegraphics[width=150pt]{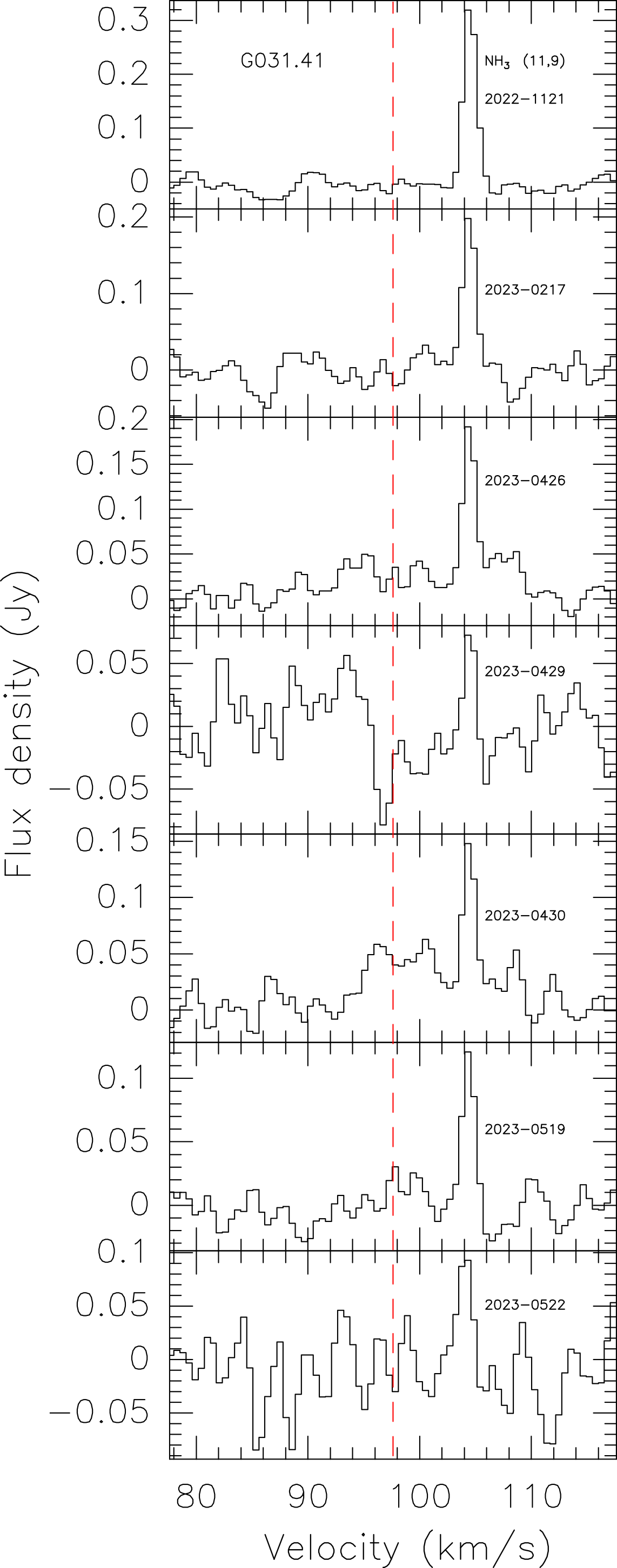}
  \caption{\label{g03141} NH$_3$ spectra towards G031.41. The dashed red lines indicate the systemic velocity.}
\end{figure}

\begin{figure}[h]
\center
    \includegraphics[width=150pt]{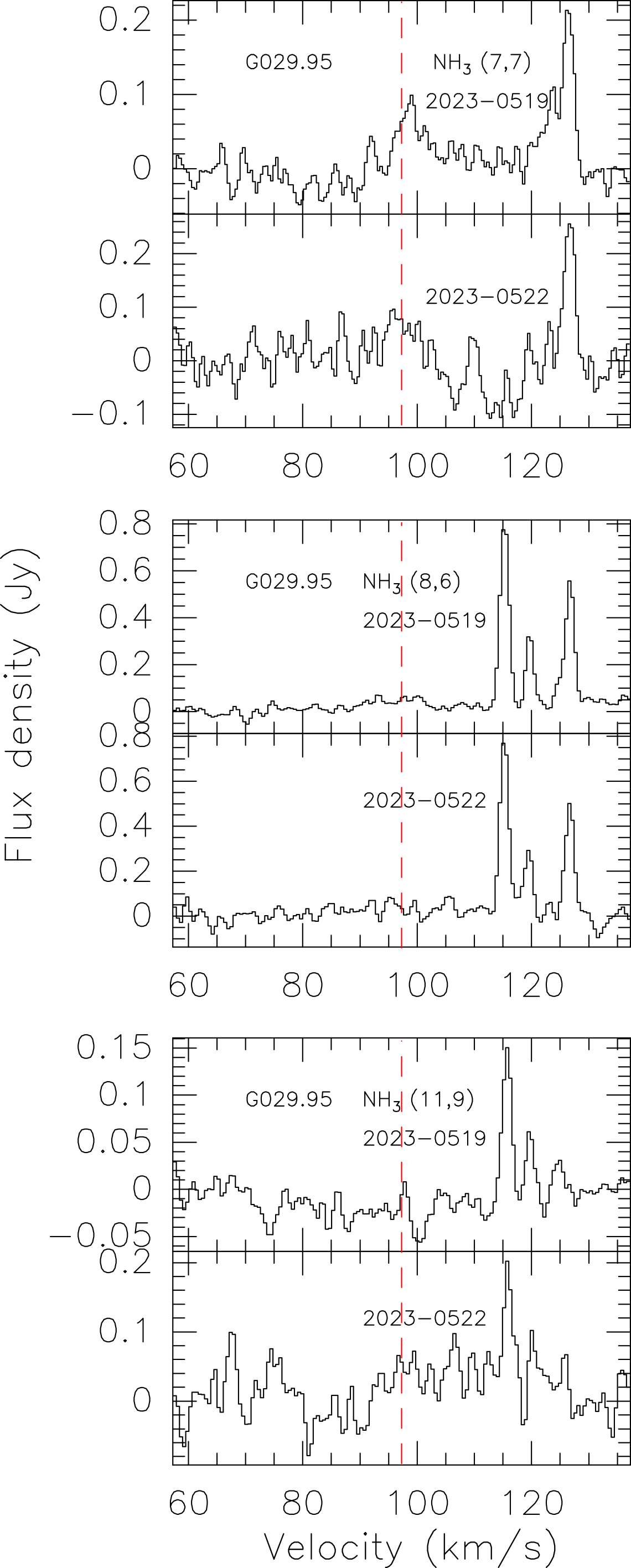}
  \caption{\label{g02995} NH$_3$ (7,7), (8,6), and (11,9) spectra towards G029.95. The dashed red lines indicate the systemic velocity.}
\end{figure}

\begin{figure*}[h]
\center
    \includegraphics[width=360pt]{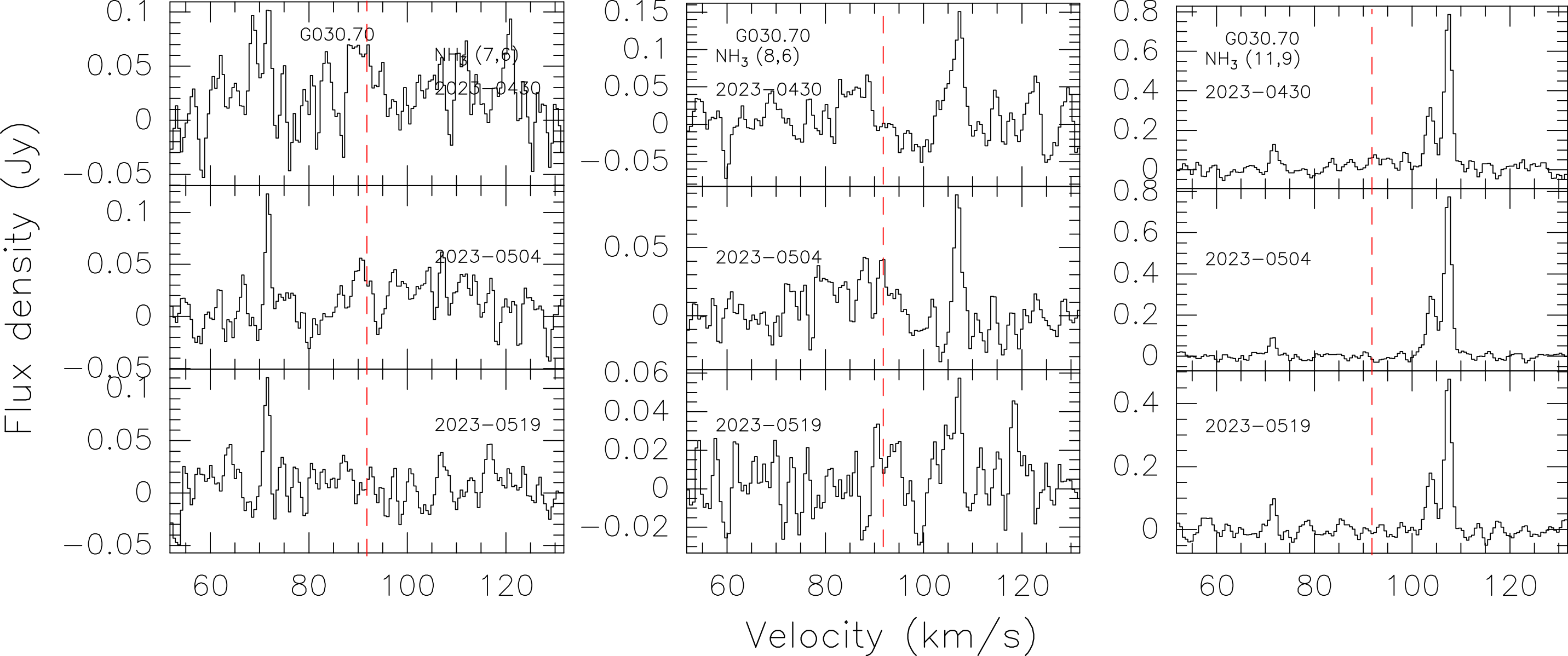}
  \caption{\label{g03070} NH$_3$ (7,6), (8,6), and (11,9) spectra towards G030.70. The dashed red lines indicate the systemic velocity.}
\end{figure*}

The NH$_3$ (11,9) transition with an energy level of 1449 K above the ground state, the highest value in our maser sample, was also detected towards G031.41 (Fig.~\ref{g03141}). Its velocity is consistent with a redshifted NH$_3$ (9,6) maser feature and its flux density decreased from November 2020 to May 2023 by 63\%. In addition, NH$_3$ (11,9) masers were also detected in G029.95-0.02 (hereafter G029.95) and G030.70-0.07 (hereafter G030.70). In G029.95, NH$_3$ (8,6) and (7,7) masers were also detected (Fig.~\ref{g02995}). The flux density of the NH$_3$ (7,7) maser increased by 55\% in three days, while the NH$_3$ (8,6) and (11,9) maser lines show no obvious variability. Toward G030.70, we also detected NH$_3$ (7,6) and (8,6) masers. Their spectra are shown in Fig.~\ref{g03070}. The NH$_3$ (7,6) maser stays constant for 20 days. During this time interval, NH$_3$ (8,6) and (11,9) masers initially showed the same trend, i.e., their flux densities are roughly the same from April 30 to May 04 but then decrease between May 04 and 19, 2023. Among the eleven sources mentioned above, G029.95 hosts both ortho- and para-NH$_3$ masers and the other ten sources only host ortho-NH$_3$ masers. 

\begin{figure*}[h]
\center
    \includegraphics[width=360pt]{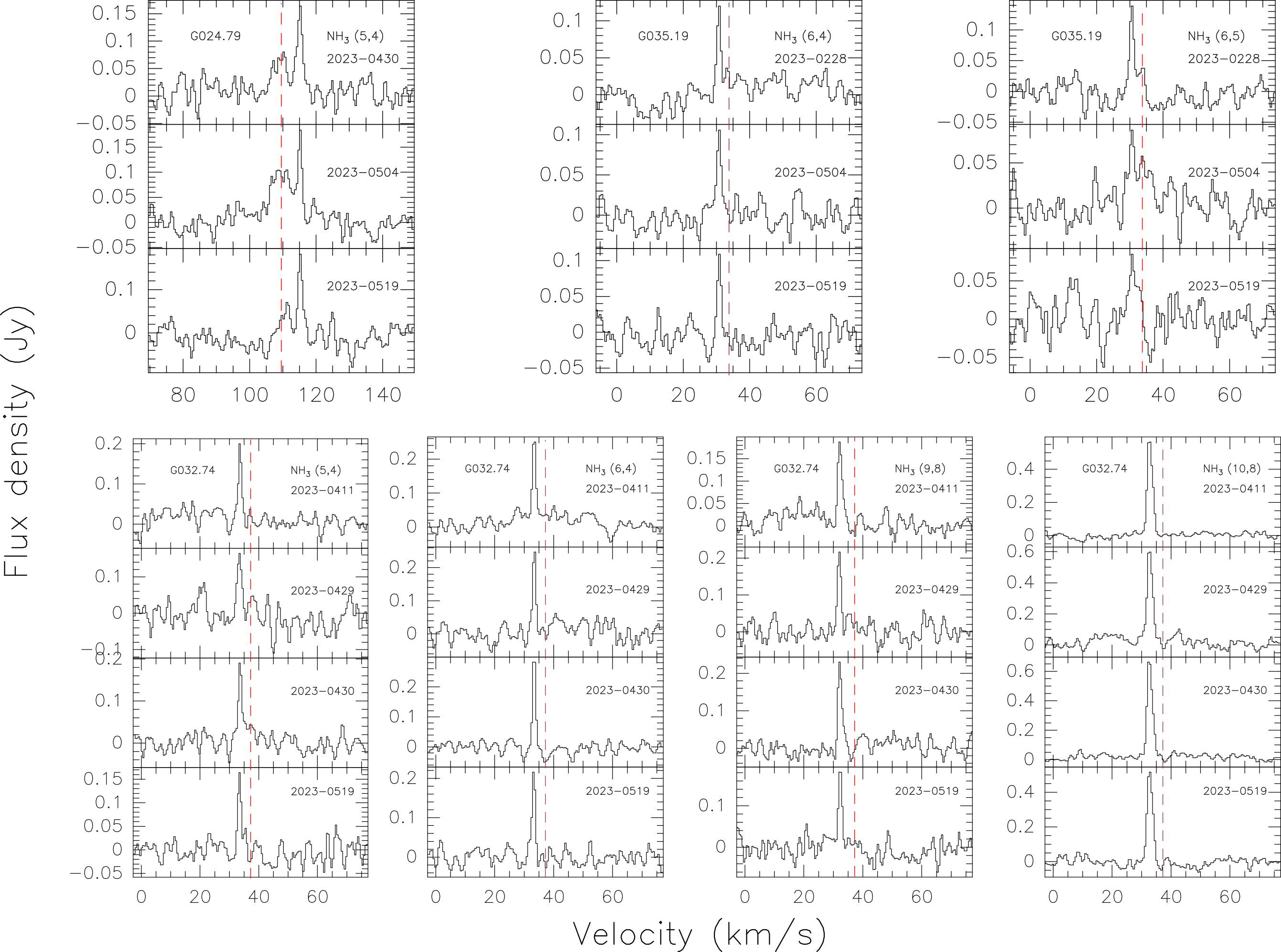}
  \caption{\label{para} NH$_3$ (5,4) spectra towards G024.79, NH$_3$ (6,4) and (6,5) spectra towards G035.19, as well as NH$_3$ (5,4), (6,4), (9,8), and (10,8) spectra towards G032.74. The dashed red lines indicate the systemic velocities of the sources.}
\end{figure*}

Three targets, G024.79+0.08 (hereafter G024.79), G032.74-0.08 (hereafter G032.74), and G035.19-0.74 (hereafter G035.19), only host para-NH$_3$ masers (Fig.~\ref{para}). Toward G024.79, we only detected an NH$_3$ (5,4) maser. Toward G032.74, four transitions were identified as masers, i.e., the NH$_3$ (5,4), (6,4), (9,8), and (10,8) lines. Two NH$_3$ maser lines, (6,4) and (6,5), were detected in G035.19. Variations of flux densities of the NH$_3$ (5,4) maser in G024.79, of the NH$_3$ (10,8) maser in G032.74, as well as from the NH$_3$ (6,5) maser in G035.19 were also observed. These variations amount to 24\% or more while the NH$_3$ (1,1) thermal emissions stay constant at the same time, thus the maser variability appears to be significant.

\begin{figure}[h]
\center
    \includegraphics[width=150pt]{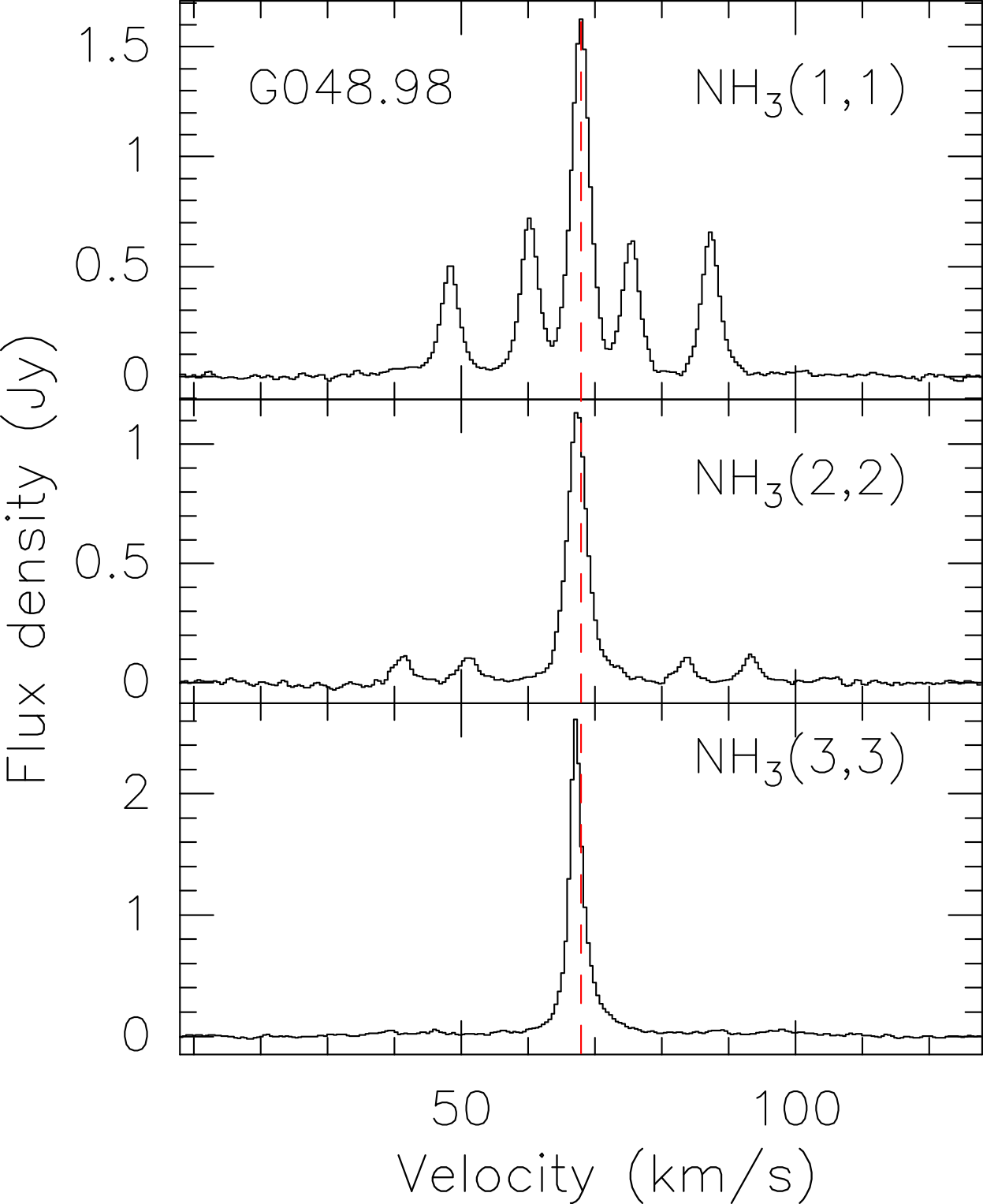}
  \caption{\label{g04898} NH$_3$ (1,1), (2,2), and (3,3) spectra towards G048.98. The dashed red lines indicate the systemic velocity.}
\end{figure}

The frequencies of the NH$_3$ (1,1), (2,2) and (3,3) transitions are within a range of only 200 MHz. The peak flux density ratios of NH$_3$ (3,3)/(1,1) and (3,3)/(2,2) towards G048.98-0.30 (hereafter G048.98) are $\sim$0.83 and $\sim$1.14, respectively, based on previous 100-m Green Bank Telescope (GBT) observations in 2010 \citep{2011MNRAS.418.1689U}, while these are $\sim$1.57 and $\sim$2.22 from our measurements. However, the ratios of NH$_3$ (1,1)/(2,2) remain consistent, which are $\sim$1.38 and $\sim$1.42, within the uncertainties due to noise in these two data sets. This indicates that the NH$_3$ (3,3) emission in G048.98 has become stronger and likely shows a maser nature. The spectra of NH$_3$ (1,1), (2,2), and (3,3) towards G048.98 are presented in Fig.~\ref{g04898}.


\begin{figure}[h]
\center
    \includegraphics[width=210pt]{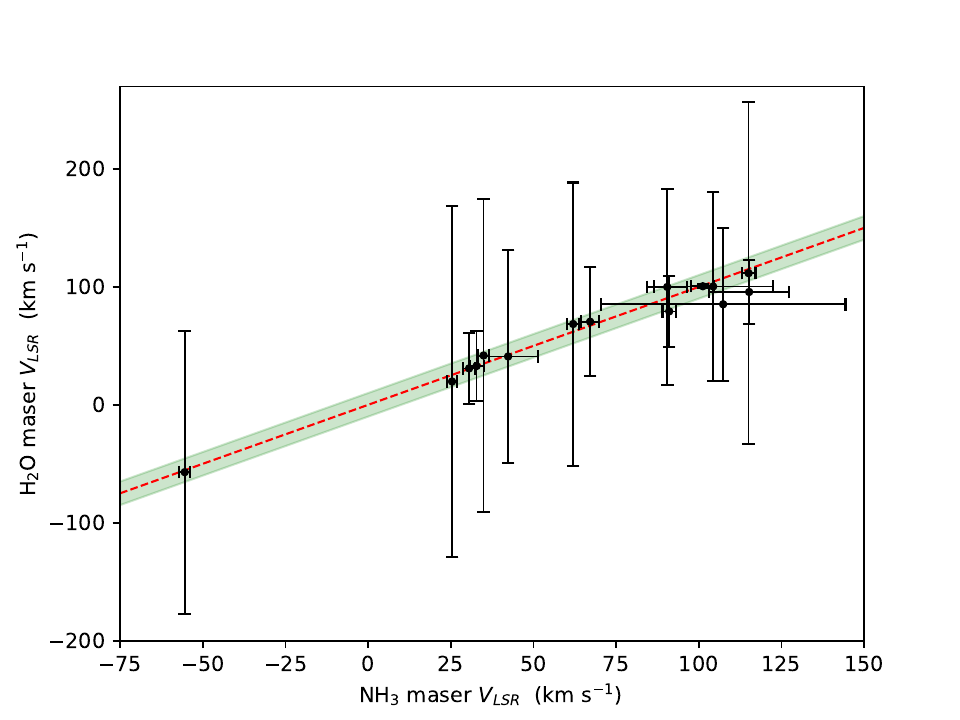}
  \caption{\label{com-h2o} Comparison of the velocity ranges of NH$_3$ and H$_2$O masers. Dots indicate the velocities of bright NH$_3$ and H$_2$O maser features. Error bars show the velocity ranges in detected NH$_3$ and H$_2$O masers. The dashed line marks locations where the velocity of NH$_3$ equals that of a water maser. The green region shows the $\pm$10 km~s$^{-1}$ zone.}
\end{figure}

In Fig~\ref{com-h2o}, we compare the velocity ranges of NH$_3$ and H$_2$O masers. The velocity range of NH$_3$ masers is always smaller than that of H$_2$O masers in each source. There are 12 objects, 80\% of our maser sample, for which the velocities of their brightest NH$_3$ maser feature are similar to that of H$_2$O masers within $\pm$10 km~s$^{-1}$. For the remaining three targets, G030.79, G029.95, and G030.70, the differences between velocities of their brightest NH$_3$ maser feature and brightest H$_2$O maser feature are large, i.e. 11~km~s$^{-1}$, 19~km~s$^{-1}$, and 22~km~s$^{-1}$, respectively.

With the single-dish observations at Effelsberg, the sizes of ammonia masers, the spatial distributions of ammonia masers and thermal emissions cannot be accurately determined. This also excludes realistic determinations of kinetic temperatures, densities, column densities, and estimates of the $\sim$10~$\mu$m infrared irradiation, potentially causing significant populations of vibrationally excited NH$_3$, right at the maser spots. So far, the NH$_3$ maser pumping processes under debate are collisional excitation, radiative excitation, and radiative excitation combined with infrared line overlap \citep[e.g.,][]{1983A&A...122..164W,1986ApJ...300L..79M,1991ApJ...378..445B,1993LNP...412..123W,2013A&A...549A..90H}. Velocity offsets with respect to systemic velocities may suggest emission associated with outflows or disks \citep{2022A&A...666L..15Y}. Higher angular resolution observations are mandatory and proposed to provide precise physical positions and sizes of the newly detected ammonia masers. This will lead to a deeper comprehension of the NH$_3$ maser phenomenon and its connection to sites of massive star formation.

\section{Summary}
\label{summary}

We report the discovery of at least 14 and likely 15 new ammonia maser sources in the Milky Way, based on our K-band line survey with the 100-meter Effelsberg telescope. Our total sample consists of 119 sources exhibiting 22\,GHz H$_2$O maser emission, thus yielding a detection rate in excess of 10\%. Fourteen of the newly detected masers are encountered in non-metastable inversion transitions and this doubles the number of non-metastable NH$_3$ masers in our Galaxy. Metastable ammonia masers are also detected in one or two sources, i.e. an NH$_3$ (7,7) maser in G029.95 and likely an NH$_3$ (3,3) maser in G048.98. Narrow line widths compared to those of ammonia $(J,K)$~=~(1,1) thermal emission, as well as variations in flux density, indicate their maser nature. All of the NH$_3$ masers in our detections have blueshifted or redshifted velocities with respect to the source systemic LSR velocities.

\begin{acknowledgements}
The authors want to thank the anonymous referee for providing useful comments, which have improved the quality of the paper. Y.T.Y. is a member of the International Max Planck Research School (IMPRS) for Astronomy and Astrophysics at the Universities of Bonn and Cologne. Y.T.Y. thanks the China Scholarship Council (CSC) for the financial support. Y.T.Y. also thanks his fianc{\'e}e, {\protect\begin{CJK*}{UTF8}{gkai}Siqi Guo (郭思祺)\protect\end{CJK*}}, for her daily and mental supports. We would like to thank the staff at the Effelsberg telescope for their help provided during the observations. 
\end{acknowledgements}

\bibliographystyle{aa}
\bibliography{ammoniamaser}

\onecolumn

\begin{appendix}
\label{appendix}

\section{Tables}
\label{appendix-tables}

\begin{center}
\begin{longtable}{lccccc}
\caption{Catalog of ammonia transitions having been identified as masers in the ISM.\label{table_nh3-masers}}\\
\hline 
\hline
Number &   Transition & $\nu$ & $E_{\rm low}/k$ & $E_{\rm up}/k$  &   Discovery   \\
       &              &       &                 &                 &    source \&    \\
       &  $(J,K)$     & (MHz) &          (K)    &          (K)    &   Reference     \\
\hline
\endfirsthead
\caption*{continued.}\\
\hline
\hline
Number &   Transition & $\nu$ & $E_{\rm low}/k$ & $E_{\rm up}/k$  &   Discovery   \\
       &              &       &                 &                 &    source \&     \\
       &  $(J,K)$     & (MHz) &          (K)    &          (K)    &   Reference     \\
\hline
\endhead
\hline
\endfoot
1  &  (1,1)   &  23694.49550  &  22.12769    &  23.26484    &  DR21, a  \\
2  &  (2,2)   &  23722.63330  &  63.30956    &  64.44806    &  Sgr B2(M), b   \\
3  &  (3,3)   &  23870.12920  &  122.39346   &  123.53904   &  W33, c   \\
4  &  (5,5)   &  24532.98870  &  294.19337   &  295.37076   &  G9.26+0.19, d   \\
5  &  (5,4)   &  22653.02200  &  342.22451   &  343.31167   &  W51-IRS2, e   \\
6  &  (5,3)   &  21285.27500  &  379.43682   &  380.45835   &  W51-IRS2, f   \\
7  &  (6,6)   &  25056.02500  &  406.85658   &  408.05908   &  NGC6334, g  \\
8  &  (6,5)   &  22732.42900  &  465.55043   &  466.64141   &  W3(OH), h  \\
9  &  (6,4)   &  20994.61700  &  513.33957   &  514.34714   &  Sgr B2(N), i  \\
10 &  (6,3)   &  19757.53800  &  550.36513   &  551.31333   &  NGC7538, j  \\
11 &  (6,2)   &  18884.69500  &  576.73689   &  577.64321   &  W51-IRS2, f  \\
12 &  (6,1)   &  18391.56200  &  592.52954   &  593.41219   &  Sgr B2(N), i  \\
13 &  (7,7)   &  25715.18200  &  537.31664   &  538.55077   &  W51-IRS2, f  \\
14 &  (7,6)   &  22924.94000  &  606.67824   &  607.77846   &  W51-IRS2, f  \\
15 &  (7,5)   &  20804.83000  &  665.02736   &  666.02583   &  W51-IRS2, e  \\
16 &  (7,4)   &  19218.46500  &  712.53838   &  713.46072   &  W51-IRS2, f  \\
17 &  (7,3)   &  18017.33700  &  749.35187   &  750.21656   &  Sgr B2(N), i  \\
18 &  (8,6)   &  20719.22100  &  834.46640   &  835.46076   &  NGC6334, k  \\
19 &  (8,5)   &  18808.50700  &  892.42835   &  893.33101   &  W51-IRS2, f  \\
20 &  (8,4)   &  17378.11000  &  939.62715   &  940.46116   &  Sgr B2(N), i  \\
21 &  (8,3)   &  16455.09900  &  976.19662   &  976.98634   &  Sgr B2(N), i  \\
22 &  (9,9)   &  27477.94300  &  851.45423   &  852.77295   &  W51-IRS2, f  \\
23 &  (9,8)   &  23657.47100  &  942.18255   &  943.31792   &  W51-IRS2, e  \\
24 &  (9,7)   &  20735.45200  &  1021.61682  &  1022.61196  &  W51-IRS2, f  \\
25 &  (9,6)   &  18499.39000  &  1089.99977  &  1090.88760  &  W51, NGC7538,  \\
   &          &               &              &              &  W49, DR21(OH), j  \\
26 &  (9,5)   &  16798.13400  &  1147.53483  &  1148.34101  &  Sgr B2(N), i  \\
27 &  (9,4)   &  15523.90000  &  1194.38934  &  1195.13437  &  Sgr B2(N), i  \\
28 &  (9,3)   &  14376.81700  &  1230.70171  &  1231.39168  &  Sgr B2(N), i  \\
29 &  (10,9)  &  24205.28700  &  1136.46581  &  1137.62748  &  W51-IRS2, f  \\
30 &  (10,8)  &  20852.52700  &  1226.43274  &  1227.43349  &  W51-IRS2, e  \\
31 &  (10,7)  &  18285.43400  &  1305.21064  &  1306.08820  &  W51-IRS2, f  \\
32 &  (10,6)  &  16319.32400  &  1373.03592  &  1373.81912  &  NGC7538, l  \\
33 &  (11,9)  &  21070.73900  &  1448.86263  &  1449.87386  &  NGC6334, k  \\
34 &  (12,12) &  31424.94300  &  1454.90272  &  1456.41088  &  W51-IRS2, f  \\
35 &  $^{15}$NH$_3$ (3,3) &  22789.42170  &  124.64182  &  125.73554  &  NGC7538, m \\
36 &  $^{15}$NH$_3$ (4,3) &  21637.89450  &  238.70035  &  239.73880  &  NGC7538, n \\
37 &  $^{15}$NH$_3$ (4,4) &  23046.01580  &  202.51921  &  203.62523  &  NGC7538, n \\
\hline
\end{longtable}
\tablefoot{The parameters of ammonia lines are taken from the Jet Propulsion Laboratory (JPL) molecular line catalog (\citealt{1998JQSRT..60..883P}, https://spec.jpl.nasa.gov/) and the Cologne Database for Molecular Spectroscopy \citep[CDMS,][]{2005JMoSt.742..215M,2016JMoSp.327...95E}. For the last column, see a \citet{1996ApJ...457L..47G}, b \citet{2018ApJ...869L..14M}, c \citet{1982A&A...110L..20W}, d \citet{1992A&A...256..618C}, e \citet{1987A&A...173..352M}, f \citet{2013A&A...549A..90H}, g \citet{2007A&A...466..989B}, h \citet{1988A&A...201..123M}, i \citet{2020ApJ...898..157M}, j \citet{1986ApJ...300L..79M}, k \citet{2007MNRAS.382L..35W}, l \citet{2012ApJ...759...76H}, m \citet{1986A&A...160L..13M}, n \citet{1991A&A...247..516S}. }
\end{center}

\begin{center}
\begin{longtable}{lccccc}
\caption{A catalog of sources hosting ammonia masers.\label{table_nh3-sources}}\\
\hline 
\hline
Number & source &   metastable  & non-metastable &   source  & Discovery   \\
       &        &               &                &    type   & Reference   \\
       &        &  $(J=K)$      &  $(J>K)$       &           &             \\
\hline
\endfirsthead
\caption*{continued.}\\
\hline
\hline
Number & source &   metastable  & non-metastable &   source  & Discovery   \\
       &        &               &                &    type   & Reference   \\
       &        &  $(J=K)$      &  $(J>K)$       &           &             \\
\hline
\endhead
\hline
\endfoot
1  &    W33               &       Y       &   N    &  SFR     &  a   \\
2  &    DR 21             &       Y       &   Y    &  SFR     &  b, c, d, e    \\
3  &    NGC7538           &       N       &   Y    &  SFR     &  c, f, g   \\
4  &    W51               &       Y       &   Y    &  SFR     &  c, h, i, j   \\
5  &    W49               &       N       &   Y    &  SFR     &  c   \\
6  &   W3(OH)             &       N       &   Y    &  SFR     &  k   \\
7  &   G9.26+0.19         &       Y       &   N    &  SFR     &  l   \\
8  &   NGC6334            &       Y       &   Y    &  SFR     &  m, n, o   \\
9  &   IRAS20126+4104     &       Y       &   N    &  SFR     &  p   \\
10 &   G05.89-0.39        &       Y       &   N    &  SFR     &  q  \\
11 &   G19.61-0.23        &       N       &   Y    &  SFR     &  r  \\
12 &   G23.33-0.30        &       Y       &   N    &  SNR     &  r, s  \\
13 &   G030.7206-00.0826  &       Y       &   N    &  SFR     &  t  \\
14 &   G20.08-0.14N       &       Y       &   N    &  SFR     &  u  \\
15 &   G35.03+0.35        &       Y       &   N    &  SFR     &  v  \\
16 &   G28.34+0.06        &       Y       &   N    &  SFR     &  w  \\
17 &   W44                &       Y       &   N    &  SNR     &  x  \\
18 &   G5.7-0.0           &       Y       &   N    &  SNR     &  x  \\
19 &   W51C               &       Y       &   N    &  SNR     &  x  \\
20 &   IC443              &       Y       &   N    &  SNR     &  x  \\
21 &   Sgr B2(M)          &       Y       &   Y    &  SFR     &  y, z \\
22 &   Sgr B2(N)          &       N       &   Y    &  SFR     &  z, z1 \\
23 &   G10.34-0.14        &       Y       &   N    &  SFR     &  z2 \\
24 &   G14.33-0.64        &       Y       &   N    &  SFR     &  z2 \\
25 &   G18.89-0.47        &       Y       &   N    &  SFR     &  z2 \\
26 &   G19.36-0.03        &       Y       &   N    &  SFR     &  z2 \\
27 &   G28.83-0.25        &       Y       &   N    &  SFR     &  z2 \\
28 &   CepA               &       N       &   Y    &  SFR     &  z3 \\
29 &   G34.26+0.15        &       N       &   Y    &  SFR     &  z3 \\
30 &   Sgr B2(NS)         &       N       &   Y    &  SFR     &  z \\
31 &   Sgr B2(S)          &       N       &   Y    &  SFR     &  z \\
32 &   G358.931-0.030     &       N       &   Y    &  SFR     &  z4 \\
33 &   G010.47+0.03       &       N       &   Y    &  SFR     & this work \\
34 &   G012.21-0.10       &       N       &   Y    &  SFR     & this work \\
35 &   G019.61-0.23       &       N       &   Y    &  SFR     & this work \\
36 &   G024.79+0.08       &       N       &   Y    &  SFR     & this work \\
37 &   G029.95-0.02  &       Y       &   Y    &  SFR     & this work \\
38 &   G030.21-0.19       &       N       &   Y    &  SFR     & this work \\
39 &   G030.60+0.18       &       N       &   Y    &  SFR     & this work \\
40 &   G030.70-0.07       &       N       &   Y    &  SFR     & this work \\
41 &   G030.79+0.20       &       N       &   Y    &  SFR     & this work \\
42 &   G031.41+0.30       &       N       &   Y    &  SFR     & this work \\
43 &   G032.74-0.08       &       N       &   Y    &  SFR     & this work \\
44 &   G035.19-0.74  &       N       &   Y    &  SFR     & this work \\
45 &   G043.79-0.13  &       N       &   Y    &  SFR     & this work \\
46 &   G048.98-0.30  &       Y       &   N    &  SFR     & this work \\
47 &   G111.53+0.76  &       N       &   Y    &  SFR     & this work \\
count &                   &       24      &   28   &          &     \\
\hline
\end{longtable}
\tablefoot{ SFR: star-forming region, SNR: supernova remnant. Y indicates a detection and N marks a non-detection. a \citet{1982A&A...110L..20W}, b \citet{1983A&A...124..322G}, c \citet{1986ApJ...300L..79M}, d \citet{1994ApJ...428L..33M}, e \citet{1996ApJ...457L..47G}, f \citet{2012ApJ...759...76H}, g \citet{2014ApJ...782...83H}, h \citet{1995ApJ...450L..63Z}, i \citet{1987A&A...173..352M}, j \citet{2013A&A...549A..90H}, k \citet{1988A&A...201..123M}, l \citet{1992A&A...256..618C}, m \citet{1995ApJ...439L...9K}, n \citet{2007A&A...466..989B}, o \citet{2007MNRAS.382L..35W}, p \citet{1999ApJ...527L.117Z}, q \citet{2008ApJ...680.1271H}, r \citet{2011MNRAS.416.1764W}, s \citet{2019ApJ...887...79H}, t \citet{2011MNRAS.418.1689U}, u \citet{2009ApJ...706.1036G}, v \citet{2011ApJ...739L..16B}, w \citet{2012ApJ...745L..30W}, x \citet{2016ApJ...826..189M}, y \citet{2018ApJ...869L..14M}, z  \citet{2022A&A...666L..15Y}, z1 \citet{2020ApJ...898..157M}, z2 \citet{2021ApJ...923..263T}, z3, \citet{2022A&A...659A...5Y}, z4 \citet{2023MNRAS.522.4728M}  }
\end{center}

\begin{center}
\begin{longtable}{lcccc}
\caption{Our sample of 119 observed sources.\label{table_sources}}\\
\hline 
\hline
Source & R.A. & Dec. &  $V_{\rm LSR}$ & Log($N_{\rm NH_3}$)  \\
 & ($h\quad m\quad s$) & ($\degr\quad \arcmin\quad \arcsec$) &  (km~s$^{-1}$) &  (cm$^{-2}$) \\
\hline
\endfirsthead
\caption{continued.}\\
\hline
\hline
Source & R.A. & Dec. &  $V_{\rm LSR}$ & Log($N_{\rm NH_3}$)  \\
 & ($h\quad m\quad s$) & ($\degr\quad \arcmin\quad \arcsec$) &  (km~s$^{-1}$) &  (cm$^{-2}$) \\
\hline
\endhead
\hline
\endfoot
G010.47+0.03      &  18  08  38.1  & $-$19  51  50 & 65.20    & 16.01 \\
G010.63-0.34      &  18  10  18.6  & $-$19  54  17 & -4.57    & 15.56 \\
G010.68-0.03      &  18  09  15.8  & $-$19  42  27 & 51.04    & 15.55 \\
G010.99-0.08      &  18  10  06.5  & $-$19  27  46 & 29.67    & 15.67 \\
G011.90-0.14      &  18  12  10.8  & $-$18  41  36 & 38.41    & 15.79 \\
G011.94-0.61      &  18  14  00.9  & $-$18  53  24 & 36.72    & 15.63 \\
G012.20-0.03 &  18  12  23.6  & $-$18  22  52 & 50.86    & 15.70 \\
G012.21-0.10      &  18  12  39.2  & $-$18  24  22 & 24.46    & 15.61 \\
G012.68-0.18      &  18  13  54.2  & $-$18  01  46 & 55.37    & 15.84 \\
G012.89+0.49      &  18  11  51.1  & $-$17  31  27 & 34.30    & 15.62 \\
G012.91-0.26      &  18  14  39.1  & $-$17  51  58 & 38.05    & 15.55 \\
G013.18+0.06      &  18  14  00.9  & $-$17  28  41 & 49.06    & 15.67 \\
G013.66-0.60      &  18  17  24.3  & $-$17  22  12 & 46.75    & 15.67 \\
G013.87+0.28      &  18  14  35.5  & $-$16  45  41 & 48.52    & 15.53 \\
G014.10+0.09      &  18  15  46.2  & $-$16  39  09 & 9.15     & 15.54 \\
G014.20-0.19      &  18  16  58.8  & $-$16  42  05 & 39.91    & 15.53 \\
G014.49-0.14      &  18  17  22.0  & $-$16  25  00 & 39.57    & 15.81 \\
G014.60+0.02 &  18  16  59.7  & $-$16  14  51 & 26.62    & 15.60 \\
G014.61+0.01      &  18  17  02.1  & $-$16  14  36 & 25.14    & 15.62 \\
G014.99-0.67 &  18  20  19.4  & $-$16  13  31 & 19.43    & 16.00 \\
G016.87-2.15 &  18  29  24.3  & $-$15  15  42 & 18.58    & 16.00 \\
G018.17-0.30      &  18  25  07.5  & $-$13  14  33 & 49.14    & 15.55 \\
G018.46-0.00      &  18  24  35.8  & $-$12  51  08 & 53.72    & 15.69 \\
G018.66+0.04 &  18  24  50.4  & $-$12  39  21 & 81.05    & 15.70 \\
G018.89-0.47      &  18  27  07.6  & $-$12  41  40 & 66.28    & 15.66 \\
G019.07-0.29 &  18  26  48.1  & $-$12  26  28 & 65.48    & 15.70 \\
G019.47+0.17      &  18  25  54.5  & $-$11  52  35 & 21.03    & 15.88 \\
G019.61-0.14      &  18  27  17.4  & $-$11  53  56 & 58.65    & 15.81 \\
G019.61-0.23      &  18  27  38.1  & $-$11  56  39 & 40.76    & 15.80 \\
G019.61-0.26      &  18  27  43.8  & $-$11  57  06 & 42.91    & 15.52 \\
G019.88-0.53 &  18  29  14.3  & $-$11  50  30 & 43.48    & 15.60 \\
G019.92-0.26      &  18  28  19.2  & $-$11  40  36 & 64.70    & 15.72 \\
G020.36-0.01      &  18  28  15.8  & $-$11  10  26 & 52.15    & 15.53 \\
G020.74-0.09 &  18  29  16.9  & $-$10  52  27 & 58.21    & 15.60 \\
G020.98+0.10      &  18  29  02.7  & $-$10  34  28 & 18.34    & 15.52 \\
G022.35+0.06 &  18  31  44.2  & $-$09  22  17 & 84.18    & 15.60 \\
G023.01-0.41      &  18  34  39.7  & $-$09  00  42 & 77.97    & 15.66 \\
G023.20+0.00      &  18  33  32.8  & $-$08  39  10 & 76.17    & 15.55 \\
G023.21-0.38      &  18  34  54.9  & $-$08  49  17 & 77.87    & 15.68 \\
G023.26+0.07 &  18  33  24.7  & $-$08  33  50 & 77.93    & 15.60 \\
G023.48+0.10      &  18  33  43.0  & $-$08  21  30 & 85.41    & 15.67 \\
G024.18+0.12 &  18  34  57.2  & $-$07  43  27 & 113.53   & 15.80 \\
G024.41+0.10      &  18  35  26.2  & $-$07  31  44 & 113.62   & 15.54 \\
G024.44-0.23      &  18  36  40.7  & $-$07  39  27 & 58.11    & 15.63 \\
G024.49-0.04      &  18  36  05.6  & $-$07  31  31 & 109.74   & 15.58 \\
G024.67-0.15      &  18  36  50.0  & $-$07  24  49 & 112.97   & 15.62 \\
G024.79+0.08      &  18  36  12.4  & $-$07  12  10 & 109.49   & 15.69 \\
G024.80+0.10 &  18  36  10.7  & $-$07  11  18 & 109.93   & 15.70 \\
G024.92+0.08      &  18  36  27.1  & $-$07  05  10 & 43.76    & 15.54 \\
G025.46-0.21 &  18  38  30.1  & $-$06  44  29 & 118.95   & 15.60 \\
G025.72+0.05 &  18  38  02.8  & $-$06  23  46 & 99.19    & 15.70 \\
G027.37-0.17      &  18  41  51.2  & $-$05  01  43 & 91.05    & 15.51 \\
G027.56+0.08      &  18  41  19.8  & $-$04  44  21 & 84.21    & 15.65 \\
G028.20-0.05      &  18  42  58.1  & $-$04  13  58 & 96.08    & 15.53 \\
G028.61+0.02 &  18  43  28.9  & $-$03  50  17 & 102.02   & 15.70 \\
G028.65+0.03      &  18  43  31.1  & $-$03  47  54 & 103.59   & 15.57 \\
G028.81+0.17      &  18  43  18.8  & $-$03  35  26 & 105.80   & 15.56 \\
G028.82+0.36      &  18  42  37.2  & $-$03  29  42 & 86.40    & 15.51 \\
G028.84+0.49      &  18  42  12.3  & $-$03  24  54 & 86.43    & 15.53 \\
G029.40-0.09      &  18  45  18.9  & $-$03  11  19 & 105.58   & 15.53 \\
G029.59-0.61 &  18  47  31.7  & $-$03  15  14 & 76.51    & 15.70 \\
G029.81+2.22 &  18  37  50.5  & $-$01  45  39 & 45.42    & 15.60 \\
G029.91-0.05      &  18  46  05.5  & $-$02  42  27 & 99.71    & 15.56 \\
G029.93-0.06      &  18  46  10.7  & $-$02  41  52 & 99.20    & 15.54 \\
G029.95-0.02 &  18  46  04.2  & $-$02  39  21 & 97.24    & 15.60 \\
G030.01-0.27      &  18  47  03.8  & $-$02  43  39 & 103.38   & 15.71 \\
G030.21-0.19      &  18  47  08.3  & $-$02  30  34 & 105.43   & 15.57 \\
G030.30+0.05 &  18  46  26.0  & $-$02  19  11 & 108.07   & 15.60 \\
G030.42-0.23      &  18  47  40.3  & $-$02  20  29 & 105.61   & 15.57 \\
G030.59-0.04 &  18  47  18.9  & $-$02  06  16 & 41.60    & 15.70 \\
G030.60+0.18      &  18  46  33.3  & $-$01  59  35 & 105.21   & 15.86 \\
G030.68-0.07      &  18  47  35.6  & $-$02  02  08 & 91.58    & 15.64 \\
G030.70-0.07      &  18  47  36.1  & $-$02  00  59 & 91.76    & 15.91 \\
G030.73-0.07      &  18  47  40.3  & $-$01  59  30 & 91.71    & 15.67 \\
G030.74-0.06      &  18  47  38.9  & $-$01  58  32 & 96.12    & 15.55 \\
G030.77-0.22      &  18  48  16.1  & $-$02  01  11 & 104.90   & 15.54 \\
G030.79+0.20      &  18  46  47.6  & $-$01  49  02 & 81.51    & 15.59 \\
G030.83-0.06      &  18  47  48.9  & $-$01  54  01 & 97.02    & 15.61 \\
G031.28+0.06      &  18  48  12.1  & $-$01  26  31 & 109.04   & 15.60 \\
G031.41+0.30      &  18  47  34.6  & $-$01  13  01 & 97.60    & 15.89 \\
G032.04+0.06 &  18  49  36.6  & $-$00  45  44 & 95.29    & 15.80 \\
G032.15+0.13      &  18  49  31.8  & $-$00  38  07 & 93.95    & 15.70 \\
G032.45+0.39      &  18  49  11.6  & $-$00  14  57 & 49.17    & 15.65 \\
G032.74-0.08      &  18  51  21.8  & $-$00  12  08 & 37.17    & 15.72 \\
G032.82-0.33 &  18  52  24.6  & $-$00  14  58 & 79.30    & 15.70 \\
G033.21-0.01      &  18  51  58.2  & $+$00  14  31 & 99.96    & 15.58 \\
G033.91+0.11 &  18  52  50.5  & $+$00  55  29 & 107.46   & 15.60 \\
G034.40+0.23 &  18  53  18.6  & $+$01  24  49 & 57.42    & 15.80 \\
G034.82+0.35 &  18  53  37.9  & $+$01  50  31 & 56.90    & 15.60 \\
G035.19-0.74 &  18  58  13.1  & $+$01  40  39 & 33.71    & 15.60 \\
G035.23-0.36      &  18  56  54.1  & $+$01  52  45 & 52.90    & 15.59 \\
G035.47+0.14 &  18  55  34.4  & $+$02  19  09 & 76.91    & 15.60 \\
G035.58-0.03 &  18  56  22.6  & $+$02  20  30 & 52.37    & 15.60 \\
G035.75+0.15      &  18  56  02.1  & $+$02  34  39 & 83.52    & 15.64 \\
G036.41+0.02 &  18  57  41.9  & $+$03  06  06 & 57.77    & 15.80 \\
G037.55+0.20 &  18  59  09.9  & $+$04  12  15 & 85.06    & 15.70 \\
G037.74-0.11      &  19  00  36.9  & $+$04  13  23 & 46.45    & 15.54 \\
G037.76-0.22      &  19  01  02.2  & $+$04  12  06 & 64.39    & 15.59 \\
G037.82+0.41      &  18  58  53.9  & $+$04  32  18 & 17.97    & 15.58 \\
G037.87-0.60 &  19  02  36.1  & $+$04  06  59 & 50.71    & 15.60 \\
G038.69-0.45      &  19  03  35.5  & $+$04  55  11 & 50.86    & 15.54 \\
G040.43+0.70      &  19  02  39.6  & $+$06  59  11 & 12.71    & 15.64 \\
G040.60-0.72 &  19  08  03.3  & $+$06  29  12 & 65.70    & 15.70 \\
G043.18-0.52 &  19  12  08.8  & $+$08  52  05 & 58.03    & 15.60 \\
G043.31-0.21 &  19  11  16.9  & $+$09  07  30 & 59.55    & 15.70 \\
G043.79-0.13 &  19  11  54.1  & $+$09  35  50 & 44.11    & 15.60 \\
G043.89-0.78 &  19  14  26.1  & $+$09  22  35 & 54.01    & 15.60 \\
G045.46+0.04 &  19  14  25.6  & $+$11  09  27 & 62.13    & 15.70 \\
G045.80-0.36      &  19  16  31.1  & $+$11  16  12 & 58.42    & 15.67 \\
G045.94-0.40      &  19  16  56.0  & $+$11  21  54 & 61.74    & 15.65 \\
G048.98-0.30 &  19  22  26.2  & $+$14  06  38 & 67.66    & 15.60 \\
G049.27-0.34      &  19  23  07.0  & $+$14  20  15 & 68.44    & 15.54 \\
G049.45-0.35 &  19  23  33.1  & $+$14  29  43 & 67.49    & 16.00 \\
G051.68+0.72      &  19  23  58.6  & $+$16  57  44 & 4.80     & 15.63 \\
G053.18+0.21      &  19  28  51.9  & $+$18  02  37 & 1.59     & 15.61 \\
G069.54-0.97 &  20  10  08.9  & $+$31  31  34 & 11.18    & 15.70 \\
G078.98+0.35 &  20  31  10.7  & $+$40  03  14 & 5.98     & 15.60 \\
G108.76-0.99 &  22  58  47.5  & $+$58  45  01 & -51.46   & 15.60 \\
G111.53+0.76 &  23  13  43.9  & $+$61  26  58 & -56.13   & 15.70 \\
\hline
\end{longtable}
\tablefoot{Column (1): source name; Columns (2) and (3): Equatorial coordinates ($J$2000); Column (4): Local Standard of Rest velocity; Column (5): NH$_3$ column densities.}
\end{center}

\begin{center}
\small
\begin{longtable}{lcccccccccc}
\caption{New ammonia masers.\label{table_masers}}\\
\hline 
\hline
Source &   transition & $\nu$ & Epoch &  Channel      & $S_\nu$  & rms    & $\int S_\nu dv$  & $V_{\rm LSR}$ & \multicolumn{2}{c}{$\Delta V_{1/2}$}  \\
       &              &       &       & spacing       &          &        &                  &               &       &    (1,1)$^*$      \\
       &  $(J,K)$     & (MHz) &       & (km s$^{-1}$) & (Jy)     &  (Jy)  & (Jy km s$^{-1}$) & \multicolumn{3}{c}{ (km s$^{-1}$)}    \\
\hline
\endfirsthead
\caption{continued.}\\
\hline
\hline
Source &   transition & $\nu$ & Epoch &  Channel      & $S_\nu$  & rms    & $\int S_\nu dv$  & $V_{\rm LSR}$ & \multicolumn{2}{c}{$\Delta V_{1/2}$}  \\
       &              &       &       & spacing       &          &        &                  &               &       &    (1,1)$^*$      \\
       &  $(J,K)$     & (MHz) &       & (km s$^{-1}$) & (Jy)     &  (Jy)  & (Jy km s$^{-1}$) & \multicolumn{3}{c}{ (km s$^{-1}$)}    \\
\hline
\endhead
\hline
\endfoot
\label{maser_fitting}
G010.47  & (9,6) & 18499.390 & 2023, May  04 & 0.62 &  0.25   & 0.015  & 0.47 $\pm$ 0.03 &  62.05 $\pm$ 0.05 &  1.79 $\pm$ 0.11 &  6.08 $\pm$ 0.48\\
         &       &           & 2023, May  22 & 0.07 &  0.17   & 0.061  & 0.08 $\pm$ 0.03 &  61.81 $\pm$ 0.04 &  0.43 $\pm$ 0.10 &  \\
         &       &           &               &      &  0.12   &        & 0.02 $\pm$ 0.01 &  62.39 $\pm$ 0.04 &  0.15 $\pm$ 0.07 &  \\
         &       &           &               &      &  0.13   &        & 0.28 $\pm$ 0.06 &  62.58 $\pm$ 0.19 &  2.00 $\pm$ 0.39 &  \\
G012.21  & (9,6) & 18499.390 & 2023, Feb. 28 & 0.62 &  0.17   & 0.013  & 0.25 $\pm$ 0.02 &  25.42 $\pm$ 0.07 &  1.40 $\pm$ 0.16 & 6.55 $\pm$ 0.48 \\
         &       &           & 2023, May  04 & 0.62 &  0.37   & 0.008  & 0.66 $\pm$ 0.01 &  25.86 $\pm$ 0.02 &  1.67 $\pm$ 0.04 &  \\
         &       &           & 2023, May  19 & 0.62 &  0.37   & 0.061  & 0.98 $\pm$ 0.15 &  25.83 $\pm$ 0.15 &  2.50 $\pm$ 0.55 &  \\
         &       &           & 2023, May  22 & 0.07 &  0.44   & 0.077  & 0.39 $\pm$ 0.03 &  26.02 $\pm$ 0.03 &  0.82 $\pm$ 0.08 &  \\
G019.612 & (9,6) & 18499.390 & 2023, Apr. 11 & 0.62 &  0.14   & 0.011  & 0.25 $\pm$ 0.02 &  34.96 $\pm$ 0.06 &  1.68 $\pm$ 0.15 & 5.93 $\pm$ 0.48 \\
         &       &           & 2023, Apr. 26 & 0.62 & $\cdots$ & 0.185  &     $\cdots$     &      $\cdots$      &     $\cdots$  &  \\
         &       &           & 2023, Apr. 30 & 0.62 &  0.27   & 0.019  & 0.59 $\pm$ 0.04 &  34.71 $\pm$ 0.07 &  2.07 $\pm$ 0.16 &  \\
         &       &           & 2023, May  19 & 0.62 &  0.27   & 0.037  & 1.60 $\pm$ 0.10 &  34.81 $\pm$ 0.15 &  3.68 $\pm$ 0.50 &  \\
         &       &           & 2023, May  22 & 0.07 &  0.18   & 0.055  & 0.05 $\pm$ 0.01 &  32.41 $\pm$ 0.03 &  0.27 $\pm$ 0.06 &  \\
         &       &           &               &      &  0.31   &        & 0.17 $\pm$ 0.03 &  33.50 $\pm$ 0.03 &  0.53 $\pm$ 0.15 &  \\
         &       &           &               &      &  0.28   &        & 0.16 $\pm$ 0.04 &  34.33 $\pm$ 0.04 &  0.52 $\pm$ 0.18 &  \\
         &       &           &               &      &  0.16   &        & 0.12 $\pm$ 0.03 &  35.28 $\pm$ 0.10 &  0.69 $\pm$ 0.21 &  \\
G024.79  & (5,4) & 22653.022 & 2023, Apr. 30 & 0.50 &  0.14   & 0.019  & 0.32 $\pm$ 0.03 & 115.10 $\pm$ 0.10 &  2.10 $\pm$ 0.25 & 3.37 $\pm$ 0.48 \\
         &       &           & 2023, May  04 & 0.50 &  0.15   & 0.018  & 0.19 $\pm$ 0.04 & 115.21 $\pm$ 0.07 &  1.19 $\pm$ 0.19 &  \\
         &       &           & 2023, May  19 & 0.50 &  0.18   & 0.022  & 0.33 $\pm$ 0.03 & 115.28 $\pm$ 0.08 &  1.70 $\pm$ 0.21 &  \\
G029.95  & (7,7) & 25715.182 & 2023, May  19 & 0.44 &  0.18   & 0.017  & 0.36 $\pm$ 0.05 & 126.56 $\pm$ 0.07 &  1.87 $\pm$ 0.18 & 3.72 $\pm$ 0.48 \\
         &       &           & 2023, May  22 & 0.44 &  0.28   & 0.030  & 0.71 $\pm$ 0.07 & 126.59 $\pm$ 0.11 &  2.36 $\pm$ 0.28 &  \\
         & (8,6) & 20719.221 & 2023, May  19 & 0.55 &  0.78   & 0.017  & 1.45 $\pm$ 0.03 & 115.26 $\pm$ 0.02 &  1.74 $\pm$ 0.03 &  \\
         &       &           &               &      &  0.30   &        & 0.45 $\pm$ 0.02 & 119.69 $\pm$ 0.04 &  1.43 $\pm$ 0.09 &  \\
         &       &           &               &      &  0.50   &        & 1.00 $\pm$ 0.03 & 126.67 $\pm$ 0.03 &  1.86 $\pm$ 0.07 &  \\
         &       &           & 2023, May  22 & 0.55 &  0.79   & 0.024  & 1.49 $\pm$ 0.05 & 115.22 $\pm$ 0.03 &  1.78 $\pm$ 0.07 &  \\
         &       &           &               &      &  0.28   &        & 0.55 $\pm$ 0.05 & 119.38 $\pm$ 0.09 &  1.85 $\pm$ 0.21 &  \\
         &       &           &               &      &  0.49   &        & 0.95 $\pm$ 0.05 & 126.66 $\pm$ 0.05 &  1.82 $\pm$ 0.11 &  \\
         & (11,9)& 21070.739 & 2023, May  19 & 0.54 &  0.16   & 0.017  & 0.29 $\pm$ 0.03 & 115.69 $\pm$ 0.08 &  1.74 $\pm$ 0.16 &  \\
         &       &           &               &      &  0.07   &        & 0.10 $\pm$ 0.02 & 119.72 $\pm$ 0.16 &  1.46 $\pm$ 0.36 &  \\
         &       &           &               &      &  0.03   &        & 0.06 $\pm$ 0.02 & 124.69 $\pm$ 0.34 &  1.66 $\pm$ 0.63 &  \\
         &       &           & 2023, May  22 & 0.54 &  0.18   & 0.035  & 0.48 $\pm$ 0.08 & 115.87 $\pm$ 0.18 &  2.51 $\pm$ 0.57 &  \\
G030.21  & (9,6) & 18499.390 & 2023, Apr. 30 & 0.62 &  0.08   & 0.015  & 0.32 $\pm$ 0.03 & 101.25 $\pm$ 0.18 &  3.51 $\pm$ 0.46 & 3.13 $\pm$ 0.05 \\
         &       &           & 2023, May  04 & 0.62 &  0.07   & 0.014  & 0.25 $\pm$ 0.04 & 101.15 $\pm$ 0.26 &  3.54 $\pm$ 0.60 &  \\
         &       &           & 2023, May  19 & 0.62 &  0.07   & 0.016  & 0.09 $\pm$ 0.02 & 100.72 $\pm$ 0.15 &  1.18 $\pm$ 0.28 &  \\
         &       &           &    average    & 0.62 &  0.07   & 0.007  & 0.14 $\pm$ 0.01 & 100.77 $\pm$ 0.09 &  1.85 $\pm$ 0.24 &  \\
G030.60  & (9,6) & 18499.390 & 2023, Apr. 30 & 0.62 &  0.11   & 0.008  & 0.25 $\pm$ 0.02 &  90.50 $\pm$ 0.08 &  2.07 $\pm$ 0.18 & 3.35 $\pm$ 0.48 \\
         &       &           &               &      &  0.06   &        & 0.08 $\pm$ 0.02 &  94.82 $\pm$ 0.13 &  1.27 $\pm$ 0.26 &  \\
         &       &           & 2023, May  04 & 0.62 &  0.09   & 0.008  & 0.19 $\pm$ 0.02 &  90.97 $\pm$ 0.10 &  1.89 $\pm$ 0.19 &  \\
         &       &           & 2023, May  19 & 0.62 &  0.20   & 0.010  & 0.53 $\pm$ 0.03 &  91.46 $\pm$ 0.06 &  2.50 $\pm$ 0.17 &  \\
G030.70  & (7,6) & 22924.940 & 2023, Apr. 30 & 0.50 &  0.09   & 0.023  & 0.18 $\pm$ 0.04 &  68.77 $\pm$ 0.21 &  1.86 $\pm$ 0.59 & 3.77 $\pm$ 0.01 \\
         &       &           &               &      &  0.10   &        & 0.16 $\pm$ 0.04 &  71.69 $\pm$ 0.18 &  1.49 $\pm$ 0.38 &  \\
         &       &           & 2023, May  04 & 0.50 &  0.12   & 0.018  & 0.14 $\pm$ 0.02 &  71.72 $\pm$ 0.08 &  1.12 $\pm$ 0.17 &  \\
         &       &           & 2023, May  19 & 0.50 &  0.12   & 0.019  & 0.17 $\pm$ 0.02 &  71.47 $\pm$ 0.10 &  1.35 $\pm$ 0.21 &  \\
         & (8,6) & 20719.221 & 2023, Apr. 30 & 0.55 &  0.12   & 0.022  & 0.38 $\pm$ 0.05 & 106.91 $\pm$ 0.20 &  2.87 $\pm$ 0.49 &  \\
         &       &           & 2023, May  04 & 0.50 &  0.10   & 0.012  & 0.17 $\pm$ 0.03 & 106.95 $\pm$ 0.13 &  1.69 $\pm$ 0.30 &  \\
         &       &           & 2023, May  19 & 0.50 &  0.05   & 0.015  & 0.06 $\pm$ 0.02 & 107.02 $\pm$ 0.25 &  1.32 $\pm$ 0.55 &  \\
         & (11,9)& 21070.739 & 2023, Apr. 30 & 0.54 &  0.12   & 0.026  & 0.23 $\pm$ 0.05 &  71.83 $\pm$ 0.18 &  1.70 $\pm$ 0.41 &  \\
         &       &           &               &      &  0.31   &        & 0.72 $\pm$ 0.05 & 103.69 $\pm$ 0.08 &  2.16 $\pm$ 0.17 &  \\
         &       &           &               &      &  0.80   &        & 1.51 $\pm$ 0.05 & 107.40 $\pm$ 0.03 &  1.78 $\pm$ 0.06 &  \\
         &       &           & 2023, May  04 & 0.54 &  0.10   & 0.012  & 0.17 $\pm$ 0.02 &  71.44 $\pm$ 0.09 &  1.65 $\pm$ 0.25 &  \\
         &       &           &               &      &  0.29   &        & 0.75 $\pm$ 0.03 & 104.02 $\pm$ 0.04 &  2.42 $\pm$ 0.11 &  \\
         &       &           &               &      &  0.81   &        & 1.44 $\pm$ 0.02 & 107.36 $\pm$ 0.01 &  1.68 $\pm$ 0.03 &  \\
         &       &           & 2023, May  19 & 0.54 &  0.11   & 0.022  & 0.15 $\pm$ 0.02 &  71.56 $\pm$ 0.11 &  1.30 $\pm$ 0.24 &  \\
         &       &           &               &      &  0.18   &        & 0.40 $\pm$ 0.03 & 103.84 $\pm$ 0.07 &  2.02 $\pm$ 0.18 &  \\
         &       &           &               &      &  0.50   &        & 0.86 $\pm$ 0.03 & 107.37 $\pm$ 0.03 &  1.61 $\pm$ 0.06 &  \\
G030.79  & (9,6) & 18499.390 & 2023, Apr. 29 & 0.62 &  0.89   & 0.019  & 1.94 $\pm$ 0.04 &  91.10 $\pm$ 0.02 &  2.04 $\pm$ 0.05 & 2.22 $\pm$ 0.48 \\
         &       &           & 2023, Apr. 30 & 0.62 &  0.98   & 0.017  & 2.03 $\pm$ 0.04 &  91.10 $\pm$ 0.02 &  1.95 $\pm$ 0.04 &  \\
         &       &           & 2023, May  19 & 0.62 &  0.33   & 0.012  & 1.16 $\pm$ 0.03 &  91.83 $\pm$ 0.05 &  3.32 $\pm$ 0.10 &  \\
G031.41  & (9,6) & 18499.390 & 2022, Nov. 21 & 0.62 &  1.86   & 0.006  & 5.32 $\pm$ 0.05 &  91.51 $\pm$ 0.01 &  2.69 $\pm$ 0.03 & 4.57 $\pm$ 0.48 \\
         &       &           &               &      &  4.51   &        & 7.56 $\pm$ 0.04 & 104.60 $\pm$ 0.01 &  1.57 $\pm$ 0.01 &  \\
         &       &           & 2022, Nov. 25 & 0.07 &  0.62   & 0.006  & 0.41 $\pm$ 0.07 &  90.53 $\pm$ 0.07 &  0.61 $\pm$ 0.07 &  \\
         &       &           &               &      &  1.47   &        & 2.15 $\pm$ 0.07 &  91.56 $\pm$ 0.07 &  1.37 $\pm$ 0.07 &  \\
         &       &           &               &      &  0.56   &        & 0.53 $\pm$ 0.07 &  93.07 $\pm$ 0.07 &  0.88 $\pm$ 0.07 &  \\
         &       &           &               &      &  5.21   &        & 4.68 $\pm$ 0.07 & 104.58 $\pm$ 0.07 &  0.84 $\pm$ 0.07 &  \\
         &       &           & 2023, Feb. 17 & 0.62 &  0.28   & 0.019  & 0.64 $\pm$ 0.08 &  88.09 $\pm$ 0.14 &  2.12 $\pm$ 0.35 &  \\
         &       &           &               &      &  1.20   &        & 2.52 $\pm$ 0.06 &  91.56 $\pm$ 0.03 &  1.98 $\pm$ 0.08 &  \\
         &       &           &               &      &  1.54   &        & 2.37 $\pm$ 0.06 & 104.43 $\pm$ 0.02 &  1.45 $\pm$ 0.04 &  \\
         &       &           & 2023, Apr. 26 & 0.62 &  0.44   & 0.008  & 1.01 $\pm$ 0.07 &  88.24 $\pm$ 0.06 &  2.15 $\pm$ 0.19 &  \\
         &       &           &               &      &  0.92   &        & 2.09 $\pm$ 0.06 &  91.92 $\pm$ 0.03 &  2.13 $\pm$ 0.08 &  \\
         &       &           &               &      &  2.18   &        & 3.54 $\pm$ 0.05 & 104.39 $\pm$ 0.01 &  1.52 $\pm$ 0.02 &  \\
         &       &           & 2023, Apr. 29 & 0.62 &  0.21   & 0.019  & 0.44 $\pm$ 0.04 &  88.14 $\pm$ 0.09 &  2.03 $\pm$ 0.23 &  \\
         &       &           &               &      &  0.45   &        & 0.95 $\pm$ 0.04 &  92.02 $\pm$ 0.03 &  2.00 $\pm$ 0.10 &  \\
         &       &           &               &      &  1.03   &        & 1.75 $\pm$ 0.03 & 104.40 $\pm$ 0.01 &  1.60 $\pm$ 0.04 &  \\
         &       &           & 2023, Apr. 29 & 0.07 &  0.48   & 0.043  & 0.59 $\pm$ 0.05 &  88.29 $\pm$ 0.07 &  1.16 $\pm$ 0.07 &  \\
         &       &           &               &      &  0.39   &        & 0.34 $\pm$ 0.05 &  91.13 $\pm$ 0.07 &  0.80 $\pm$ 0.07 &  \\
         &       &           &               &      &  0.97   &        & 1.23 $\pm$ 0.05 &  92.17 $\pm$ 0.07 &  1.19 $\pm$ 0.07 &  \\
         &       &           &               &      &  4.53   &        & 3.05 $\pm$ 0.02 & 104.39 $\pm$ 0.07 &  0.63 $\pm$ 0.07 &  \\
         &       &           &               &      &  0.24   &        & 0.35 $\pm$ 0.03 & 106.43 $\pm$ 0.07 &  1.38 $\pm$ 0.07 &  \\
         &       &           & 2023, Apr. 30 & 0.62 &  0.33   & 0.010  & 0.84 $\pm$ 0.06 &  88.25 $\pm$ 0.08 &  2.37 $\pm$ 0.24 &  \\
         &       &           &               &      &  0.69   &        & 1.59 $\pm$ 0.06 &  91.98 $\pm$ 0.04 &  2.14 $\pm$ 0.10 &  \\
         &       &           &               &      &  1.62   &        & 2.78 $\pm$ 0.05 & 104.41 $\pm$ 0.01 &  1.62 $\pm$ 0.03 &  \\
         &       &           & 2023, May  19 & 0.62 &  0.11   & 0.009  & 0.30 $\pm$ 0.20 &  88.91 $\pm$ 0.62 &  2.59 $\pm$ 0.62 &  \\
         &       &           &               &      &  1.03   &        & 2.07 $\pm$ 0.20 &  91.14 $\pm$ 0.62 &  1.88 $\pm$ 0.62 &  \\
         &       &           &               &      &  1.80   &        & 2.88 $\pm$ 0.20 & 104.40 $\pm$ 0.62 &  1.50 $\pm$ 0.62 &  \\
         &       &           & 2023, May  22 & 0.62 &  0.11   & 0.019  & 0.38 $\pm$ 0.19 &  88.10 $\pm$ 0.62 &  3.22 $\pm$ 0.62 &  \\
         &       &           &               &      &  0.86   &        & 2.14 $\pm$ 0.19 &  92.11 $\pm$ 0.62 &  2.32 $\pm$ 0.62 &  \\
         &       &           &               &      &  1.71   &        & 2.88 $\pm$ 0.19 & 104.40 $\pm$ 0.62 &  1.58 $\pm$ 0.62 &  \\
         & (11,9)& 21070.739 & 2022, Nov. 21 & 0.54 &  0.35   & 0.012  & 0.52 $\pm$ 0.02 & 104.52 $\pm$ 0.02 &  1.41 $\pm$ 0.06 &  \\
         &       &           & 2023, Feb. 17 & 0.54 &  0.21   & 0.019  & 0.31 $\pm$ 0.02 & 104.44 $\pm$ 0.05 &  1.41 $\pm$ 0.13 &  \\
         &       &           & 2023, Apr. 26 & 0.54 &  0.18   & 0.012  & 0.36 $\pm$ 0.05 & 104.49 $\pm$ 0.10 &  1.84 $\pm$ 0.34 &  \\
         &       &           & 2023, Apr. 29 & 0.54 & $\cdots$ & 0.021  &     $\cdots$   &      $\cdots$      &     $\cdots$    &    \\
         &       &           & 2023, Apr. 30 & 0.54 &  0.14   & 0.015  & 0.23 $\pm$ 0.02 & 104.46 $\pm$ 0.07 &  1.51 $\pm$ 0.20 &  \\
         &       &           & 2023, May  19 & 0.54 &  0.12   & 0.012  & 0.15 $\pm$ 0.02 & 104.37 $\pm$ 0.07 &  1.20 $\pm$ 0.14 &  \\
         &       &           & 2023, May  22 & 0.54 &  0.13   & 0.038  & 0.39 $\pm$ 0.11 & 103.91 $\pm$ 0.32 &  2.94 $\pm$ 1.27 &  \\
G032.74  & (5,4) & 22653.022 & 2023, Apr. 11 & 0.50 &  0.18   & 0.017  & 0.25 $\pm$ 0.02 &  33.61 $\pm$ 0.05 &  1.30 $\pm$ 0.14 & 4.21 $\pm$ 0.48 \\
         &       &           & 2023, Apr. 29 & 0.50 &  0.18   & 0.025  & 0.37 $\pm$ 0.05 &  33.43 $\pm$ 0.13 &  1.98 $\pm$ 0.29 &  \\
         &       &           & 2023, Apr. 30 & 0.50 &  0.18   & 0.018  & 0.34 $\pm$ 0.03 &  33.69 $\pm$ 0.08 &  1.80 $\pm$ 0.27 &  \\
         &       &           & 2023, May  19 & 0.50 &  0.17   & 0.020  & 0.25 $\pm$ 0.03 &  33.45 $\pm$ 0.07 &  1.39 $\pm$ 0.19 &  \\
         & (6,4) & 20994.617 & 2023, Apr. 11 & 0.54 &  0.26   & 0.014  & 0.43 $\pm$ 0.03 &  33.37 $\pm$ 0.06 &  1.56 $\pm$ 0.16 &  \\
         &       &           & 2023, Apr. 29 & 0.54 &  0.27   & 0.021  & 0.38 $\pm$ 0.04 &  33.39 $\pm$ 0.07 &  1.34 $\pm$ 0.18 &  \\
         &       &           & 2023, Apr. 30 & 0.54 &  0.33   & 0.012  & 0.41 $\pm$ 0.02 &  33.36 $\pm$ 0.03 &  1.17 $\pm$ 0.07 &  \\
         &       &           & 2023, May  19 & 0.54 &  0.22   & 0.019  & 0.30 $\pm$ 0.02 &  33.19 $\pm$ 0.05 &  1.28 $\pm$ 0.12 &  \\
         & (9,8) & 23657.471 & 2023, Apr. 11 & 0.48 &  0.18   & 0.020  & 0.32 $\pm$ 0.03 &  32.27 $\pm$ 0.07 &  1.70 $\pm$ 0.15 &  \\
         &       &           & 2023, Apr. 29 & 0.48 &  0.22   & 0.023  & 0.33 $\pm$ 0.03 &  32.01 $\pm$ 0.06 &  1.41 $\pm$ 0.14 &  \\
         &       &           & 2023, Apr. 30 & 0.48 &  0.23   & 0.016  & 0.39 $\pm$ 0.03 &  32.26 $\pm$ 0.06 &  1.58 $\pm$ 0.13 &  \\
         &       &           & 2023, May  19 & 0.48 &  0.20   & 0.019  & 0.32 $\pm$ 0.03 &  32.33 $\pm$ 0.06 &  1.46 $\pm$ 0.15 &  \\
         & (10,8)& 20852.527 & 2023, Apr. 11 & 0.55 &  0.58   & 0.015  & 1.32 $\pm$ 0.02 &  32.82 $\pm$ 0.02 &  2.13 $\pm$ 0.04 &  \\
         &       &           & 2023, Apr. 29 & 0.55 &  0.60   & 0.033  & 1.41 $\pm$ 0.06 &  32.79 $\pm$ 0.04 &  2.22 $\pm$ 0.11 &  \\
         &       &           & 2023, Apr. 30 & 0.55 &  0.66   & 0.019  & 1.47 $\pm$ 0.03 &  32.77 $\pm$ 0.02 &  2.11 $\pm$ 0.05 &  \\
         &       &           & 2023, May  19 & 0.55 &  0.53   & 0.023  & 1.21 $\pm$ 0.04 &  32.82 $\pm$ 0.04 &  2.16 $\pm$ 0.08 &  \\
G035.19  & (6,4) & 20994.617 & 2023, Feb. 28 & 0.54 &  0.12   & 0.013  & 0.22 $\pm$ 0.03 &  30.75 $\pm$ 0.10 &  1.73 $\pm$ 0.31 & 2.88 $\pm$ 0.48 \\
         &       &           & 2023, May  04 & 0.54 &  0.10   & 0.011  & 0.22 $\pm$ 0.03 &  30.78 $\pm$ 0.11 &  2.07 $\pm$ 0.35 &  \\
         &       &           & 2023, May  19 & 0.54 &  0.12   & 0.017  & 0.17 $\pm$ 0.02 &  30.85 $\pm$ 0.08 &  1.39 $\pm$ 0.19 &  \\
         & (6,5) & 22732.429 & 2023, Feb. 28 & 0.50 &  0.13   & 0.017  & 0.24 $\pm$ 0.03 &  30.57 $\pm$ 0.08 &  1.71 $\pm$ 0.25 &  \\
         &       &           & 2023, May  04 & 0.50 &  0.07   & 0.013  & 0.17 $\pm$ 0.07 &  30.32 $\pm$ 0.21 &  2.41 $\pm$ 0.53 &  \\
         &       &           & 2023, May  19 & 0.50 &  0.06   & 0.019  & 0.21 $\pm$ 0.04 &  31.01 $\pm$ 0.30 &  3.04 $\pm$ 0.59 &  \\
G043.79  & (9,6) & 18499.390 & 2023, Feb. 17 & 0.62 &  0.15   & 0.014  & 0.39 $\pm$ 0.05 &  35.49 $\pm$ 0.14 &  2.39 $\pm$ 0.43 & 4.47 $\pm$ 0.12 \\
         &       &           &               &      &  1.21   &        & 2.64 $\pm$ 0.04 &  42.37 $\pm$ 0.02 &  2.04 $\pm$ 0.04 &  \\
         &       &           & 2023, Apr. 29 & 0.62 &  0.09   & 0.013  & 0.16 $\pm$ 0.02 &  35.36 $\pm$ 0.11 &  1.65 $\pm$ 0.23 &  \\
         &       &           &               &      &  0.51   &        & 1.19 $\pm$ 0.02 &  42.33 $\pm$ 0.02 &  2.20 $\pm$ 0.05 &  \\
G111.53  & (9,6) & 18499.390 & 2023, Apr. 30 & 0.62 &  1.16   & 0.011  & 2.03 $\pm$ 0.02 & -55.43 $\pm$ 0.01 &  1.65 $\pm$ 0.02 & 4.18 $\pm$ 0.48 \\
         &       &           & 2023, May  19 & 0.62 &  0.91   & 0.012  & 1.62 $\pm$ 0.02 & -55.45 $\pm$ 0.01 &  1.67 $\pm$ 0.03 &  \\
         &       &           & 2023, May  22 & 0.07 &  0.36   & 0.040  & 0.24 $\pm$ 0.01 & -55.43 $\pm$ 0.02 &  0.63 $\pm$ 0.05 &  \\
G048.98  & (3,3) & 23870.100 & 2022, Nov. 21 & 0.48 &  2.40   & 0.015  & 6.93 $\pm$ 0.08 &  67.17 $\pm$ 0.01 &  2.71 $\pm$ 0.04 &  \\
         & (2,2) &           &               &      &  1.08   & 0.015  & 4.33 $\pm$ 0.04 &  67.35 $\pm$ 0.01 &  3.75 $\pm$ 0.04 &  \\
         & (1,1) &           &               &      &  1.53   & 0.009  & 5.89 $\pm$ 0.14 &  67.87 $\pm$ 0.48 &  3.61 $\pm$ 0.48 & 3.05 $\pm$ 0.4 \\
\hline
\end{longtable}
\tablefoot{ $^*$ The intrinsic line widths of ammonia (J, K) = (1,1) thermal emission.}
\end{center}

%
\section{Figures}
\label{appendix-figures}

\begin{figure*}[h]
\center
    \includegraphics[width=400pt]{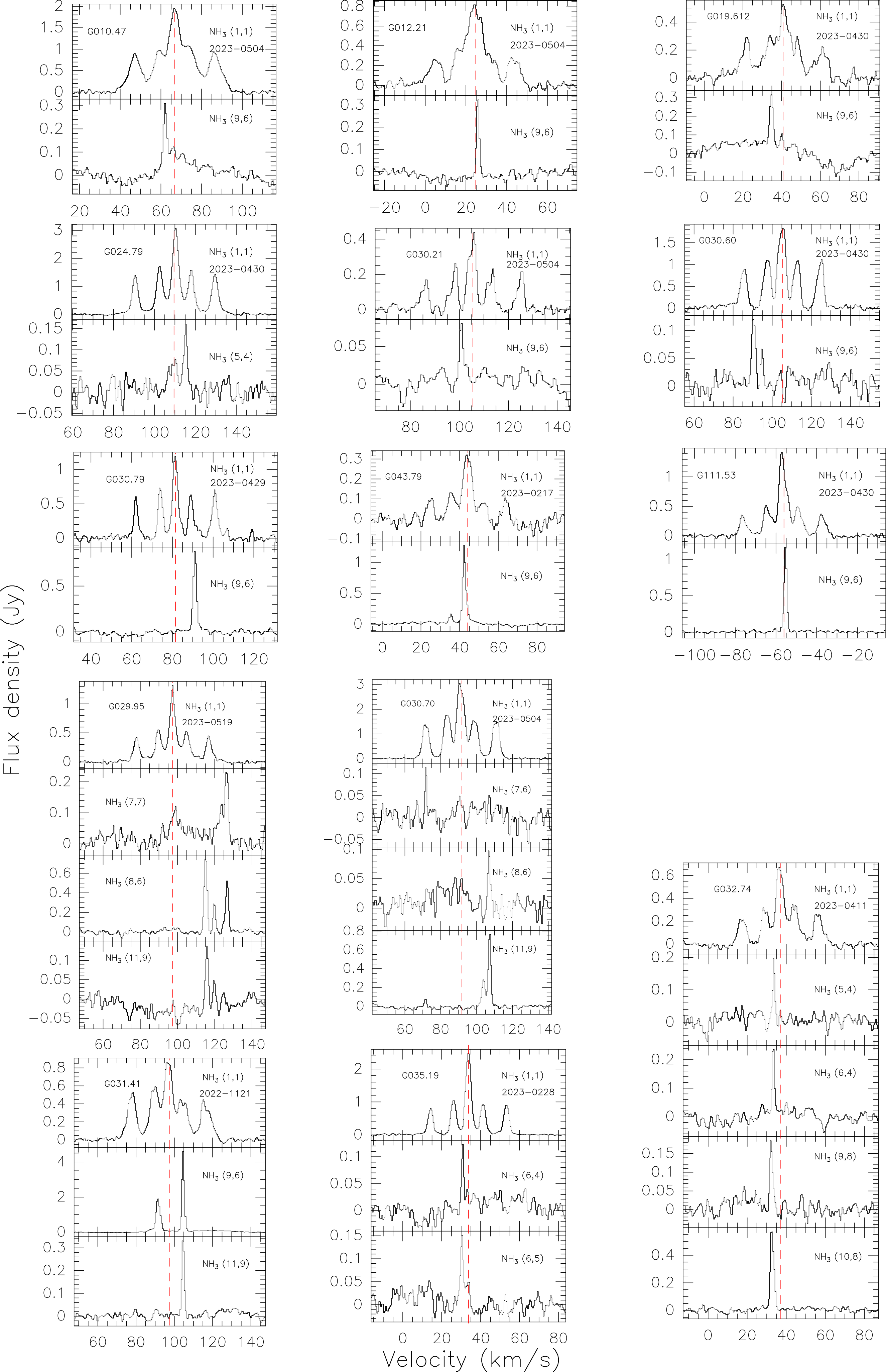}
  \caption{\label{all-11} Line profiles of NH$_3$ (1,1) thermal emission and maser transitions in 14 sources. The dashed red lines indicate the systemic velocities of the sources.}
\end{figure*}

\end{appendix}

\end{document}